\documentclass[12pt]{article}
\usepackage{amsmath,amssymb,amsfonts,amsthm,graphicx,slashed,color, epsf, dsfont, dcolumn,yfonts}
\usepackage[colorlinks=true,linkcolor=blue,linktoc=all]{hyper ref}

\def\leigh{Robert G. Leigh}
\def\uiucaddress{\small Department of Physics, University of Illinois, 1110 W. Green St., 
Urbana IL 61801-3080, U.S.A. }
\def\title{\Large {Torsional Anomalies, Hall Viscosity, and Bulk-boundary Correspondence in Topological States
}}

\parskip=\baselineskip
\renewcommand{\baselinestretch}{1.2}
\newcommand{\cDsl}{{{\cal D}\kern-.65em /\,}}
\newcommand{\cDslsm}{{{\cal D}\kern-.5em /\,}}
\newcommand{\nabslsm}{\nabla\kern-.55em /}
\newcommand{\pasl}{\pa\kern-.55em /}
\newcommand{\psl}{p\kern-.45em /}
\newcommand{\Dsl}{D\kern-.65em /}
\newcommand{\Asl}{A\kern-.55em /}
\newcommand{\nabsl}{\nabla\kern-.65em /\kern+.2em}
\newcommand{\qsl}{q\kern-.5em /}
\newcommand{\ksl}{k\kern-.5em /}
\newcommand{\rsl}{r\kern-.5em /}
\newcommand{\lcw}{\mathring{\omega}}
\newcommand{\lcR}{\mathring{R}}
\newcommand{\torcplg}{{q_T}} 

\newcommand{\cDLC}{{\stackrel{\circ}{{\cal D}}}}
\newcommand{\cDslLC}{{\stackrel{\circ}{\cDsl}}}
\newcommand{\cDslLCsq}{{\stackrel{\circ}{\cDsl^{\kern2pt 2}}}}
\newcommand{\RLC}{{\stackrel{\circ}{R}}}

\newcommand\cc[1]{#1^{^{\kern-6pt \circ}}\kern2pt}
 
\newcommand{\msD}{\mathcal{D}}
\newcommand{\e}{\underline{e}}
\newcommand{\con}{C}

\newcommand{\pa}{\partial}

\newcommand{\beq}{\begin{equation}}
\newcommand{\eeq}{\end{equation}}
\newcommand{\beqn}{\begin{eqnarray}}
\newcommand{\eeqn}{\end{eqnarray}}

\def\dalemb#1#2{{\vbox{\hrule height .#2pt
\hbox{\vrule width.#2pt height#1pt \kern#1pt
\vrule width.#2pt}
\hrule height.#2pt}}}

\begin{document}

\begin{center}
\title
\end{center}
\vskip 2 cm
\centerline{{\bf 
Taylor L. Hughes, \leigh, Onkar Parrikar}}

\vspace{.5cm}
\centerline{$^1$\it \uiucaddress}
\vspace{2cm}

\begin{abstract}
We study the transport properties of topological insulators, encoding them in a generating functional of gauge and gravitational sources. Much of our focus is on the simple example of a free massive Dirac fermion, the so-called Chern insulator, especially in 2+1 dimensions. In such cases, when parity and time-reversal symmetry are broken, it is necessary to consider the gravitational sources to include a frame and an independent spin connection with torsion. In 2+1 dimensions, the simplest parity-odd response is the Hall viscosity. We compute the Hall viscosity of the Chern insulator using a careful regularization scheme, and find that although the Hall viscosity is generally divergent, the difference in Hall viscosities of distinct topological phases is well-defined and determined by the mass gap. Furthermore, on a 1+1-dimensional edge between topological phases, the jump in the Hall viscosity across the interface is encoded, through familiar anomaly inflow mechanisms, in the structure of anomalies. In particular, we find new torsional contributions to the covariant diffeomorphism anomaly in 1+1 dimensions. Including parity-even contributions, we find that the renormalized generating functionals of the two topological phases differ by a chiral gravity action with a negative cosmological constant. This (non-dynamical) chiral gravity action and the corresponding physics of the interface theory is reminiscent of well-known properties of dynamical holographic gravitational systems. Finally, we consider some properties of spectral flow of the edge theory driven by torsional dislocations.
\end{abstract}

\pagebreak

\parskip= 2pt
\renewcommand{\baselinestretch}{.2}
\tableofcontents

\newpage

\parskip=10pt
\renewcommand{\baselinestretch}{1.2}

\section{Introduction}
Quantum field theory anomalies imply that symmetries that were present in the classical Lagrangian are broken due to quantum effects. While at one time they might have been thought of as a sickness of certain field theories, anomalies lie at the heart of some of the most fundamental physical phenomena in real materials. The canonical example is the integer quantum Hall effect (IQHE) where a 2+1-dimensional electron gas in a large, uniform magnetic field exhibits a Hall conductance which is quantized in units of $e^2/h$ when the chemical potential lies in a Landau level gap (and has been measured to be quantized up to 10 significant digits). The precise quantization arises from the connection between the Hall conductance and a topological invariant of 2+1-d electron systems called the first Chern number $C_1.$ Since $C_1$ is a topological quantity which is determined by the ground state, it is not affected when the system is perturbed continuously, and is insensitive to the microscopic details of the sample as long as the bulk energy-gap is not destroyed. Thus, response coefficients that are determined by topological invariants are the most universal features of gapped systems. 

For all understood topological response coefficients there is a complementary way to view the quantization by studying the properties of the gapless, fermionic modes that lie on the boundary of the system. There is a deep connection between topological transport in the bulk of a gapped material (say in 2+1-d) and field theory anomalies that are present for the (say 1+1-d) gapless boundary states\cite{callanharvey,kao1996}. The connection between anomalous currents, topology, and index theorems underlies some of the most beautiful transport phenomena that have been predicted, and in some cases observed in real materials. For the IQHE this bulk-boundary correspondence connects the bulk Hall transport to the spectral flow of the boundary chiral modes due to the chiral anomaly. The edge anomaly provides a complementary picture of  the origin of the Hall conductance quantization which is commonly known as Laughlin's gauge argument (though it was not originally written in terms of anomalies)\cite{laughlin1981}. 

\begin{figure}[t]
\begin{center}
\includegraphics[width=4.85in]{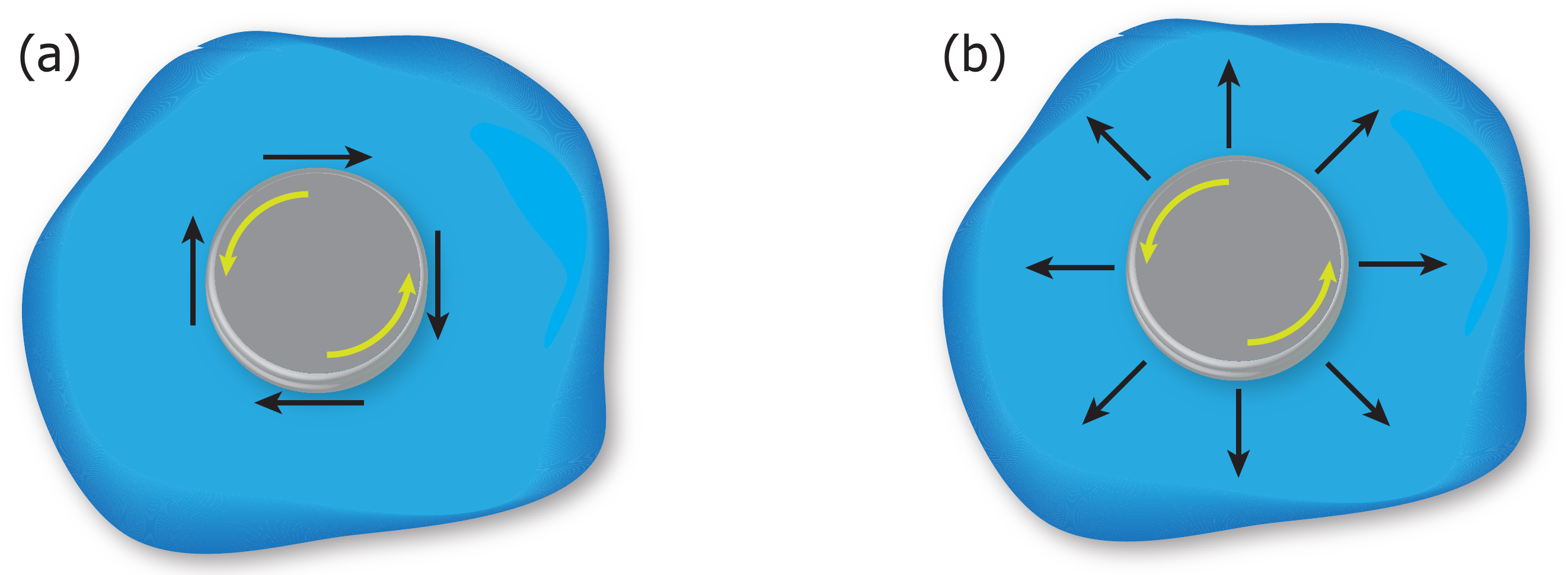}
\caption{Fluid mechanics illustration of the viscous forces. A counter-clockwise rotating solid cylinder immersed in 2d liquid droplet with (a) non-zero shear viscosity (b) non-zero dissipationless viscosity. Note that the resulting forces (arrows outside cylinder) are tangent and perpendicular to the cylinder motion (arrows inside cylinder) respectively. The shear viscosity impedes the cylinder while the dissipationless viscosity pushes fluid toward or away from the cylinder depending on the rotation direction.} \label{fig:droplet}
\end{center}
\end{figure}

While most anomalies connected with charge and spin currents are well understood, the anomalous thermal, and visco-elastic responses (VE) are not. The thermal and VE responses lie at the intersection between geometry, topology, and quantum field theory as they are usually represented as topological phenomena associated to geometric deformations of a field theory.
One example of such a novel effect is a dissipationless, electronic viscosity response in the 2+1-d topological Chern insulator with broken time-reversal symmetry\cite{avron1995,haldane1988,hughes2011}. While the ordinary shear viscosity generates a frictional force \emph{tangent} to fluid motion, the dissipationless viscosity produces a \emph{perpendicular} force (see Figure \ref{fig:droplet})\cite{avron1995,avron1998}. 
This viscosity is not clearly understood except in some special cases including the integer and fractional QHE with rotation\cite{read2009,read2011} and translation invariance\cite{haldane2009A}, and chiral superconductors\cite{read2009,bradlyn2012}. However, all of these models share the feature that they are Galilean invariant, and in relativistic systems, or lattice models with broken continuous translation symmetry, it is not clear if the topological viscosity is quantized, or even well-defined (for the lattice case)\cite{hughes2011}. This is unusual as one would expect that it should be quantized like \emph{all} of the other examples of topological response coefficients, such as the quantized Hall conductance (which is simultaneously present in the 2+1-d Chern insulator phase)\cite{qi2008b}.  

In this article we will address these issues by constructing an explicit bulk-boundary correspondence which allows us to understand the anomaly mechanism associated to the topological viscosity. The interplay of the \emph{topological} response with the \emph{geometric} deformations of the system makes this problem more subtle than previous known examples of topological responses, because, while topology does not care about the details of a shape, geometry does. The bulk-boundary correspondence for the viscosity response is completely unknown and, as we will indicate below, must have a different physical origin than the, say, chiral anomaly. The model we will focus on for most of this work is the massive Dirac model. This model represents the low-energy physics of topological insulators in various dimensions, and with various symmetries\cite{qi2008}. This model responds quite differently to geometric perturbations than typical non-relativistic electrons (\emph{i.e.} systems with small spin-orbit coupling). To illustrate the underlying premise, we first note that conventional non-relativistic electrons in a crystal are described by the Schrodinger equation at low-energy, and are only elastically influenced by the stretching of bonds that is captured by the strain tensor\cite{landauElasticity}. However, spin-orbit coupled electrons described, for example, by the Dirac equation at low-energy, are also aware of the local orbital orientation, which is not contained in the strain tensor. Instead the Dirac model couples to geometric perturbations via a local ``frame field" that we will introduce below. This additional sensitivity generates physical responses to shearing, twisting, and compressing/stretching that are not  found in weakly spin-orbit coupled systems. These phenomena are the focus of our work and are connected with the idea of geometric \emph{torsion} as we will discuss.

Our paper is organized as follows: In section \ref{section 2}, we introduce some basic concepts of geometry and elasticity, with a special emphasis on torsion, from a condensed matter perspective that are relevant to our later discussions. We will follow this up with a more mathematically precise description in section \ref{section 3}, from the point of view of general relativity and high energy physics. In section \ref{section 4}, we introduce some basic aspects of fermions in the presence of background gauge and gravitational fields, again focussing on the role of torsion. Through sections \ref{section 3} and \ref{section 4}, we will also set up notation that will be used in the rest of the paper. In section \ref{section 5}, we calculate and carefully regularize the Hall viscosity for the Dirac model in 2+1 dimensions, working about a flat background. In section \ref{sec: chiral gravity}, we will then proceed to compute the full effective action in the large mass limit on a generic background, and show the emergence of the chiral gravity action in the non-trivial topological insulator phase. One of the main ideas in sections \ref{section 5} and \ref{sec: chiral gravity} will be, that the \emph{differences} in transport coefficients between different phases are physically meaningful. We will explore this further in sections \ref{section 7} and \ref{section 8}, where we study the parity-odd transport properties in the context of anomalies due to chiral edge states localized on the interface separating a non-trivial phase from a trivial one. Finally, we will study Hall viscosity from the point of view of the interface Hamiltonian spectral flow in section \ref{sec:spectralflow}, and discuss possible mechanisms for the corresponding bulk-boundary momentum transfer.

\section{Informal Preliminaries} \label{section 2}
\begin{figure}[t]
\begin{center}
\includegraphics[width=6.45in]{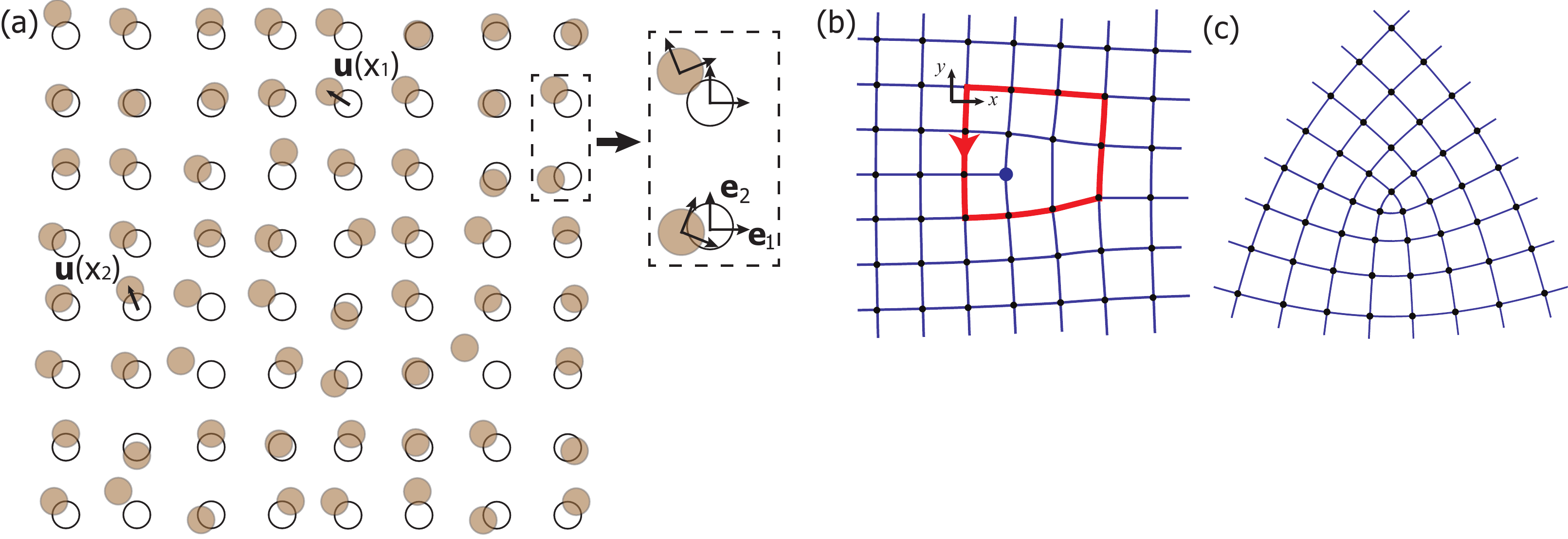}
\caption{(a) Reference state (hollow circles) and displaced state (solid circles) for an elastic medium. Displacement vectors for each site $n$ are denoted by $\mathbf{u}(x_n).$ Zoom-in shows frame field vectors $\mathbf{e}_1,\mathbf{e}_2$ in the reference state (aligned to crystal $x,y$-axes) and the displaced state (rotated with respect to crystal axes). (b) Edge dislocation representing the fundamental \emph{torsion} lattice defect. An electron traveling the thick line surrounding the dislocation will be translated with respect to the same path in the reference state that does not enclose a dislocation. The Burgers vector is in the $y$-direction. (c) Disclination represented by a single triangular plaquette in a square lattice crystal. Gives rise to \emph{curvature} \emph{i.e.} objects that travel around a disclination are rotated with respect to the reference-state path.} \label{fig:elastic}
\end{center}
\end{figure}
Before we move on to a more precise description with which high-energy theorists will be more comfortable, we try to informally introduce the necessary background material for a condensed-matter audience using the language of elasticity theory.
Conventional elasticity theory is one of the foundational underpinnings of solid state physics as it contains within it the physics of the lattice structure, including, for example, phonon fluctuations away from the ordered reference state. At a given time, one characterizes an elastic medium via a displacement field $\mathbf{u}(x_n)$ which gives the vector displacement of a lattice site $n,$ away from the position $x_n$ of a given reference state (note that we will take the continuum limit where $n$ becomes a continuous label and thus $x_n$ becomes a continuous coordinate yielding a \emph{field} $\mathbf{u}(x)$) . If every lattice point is displaced by the same amount then the crystal has just been globally translated and does not feel any internal stress. However, if the displacements of lattice sites are not identical, the material will respond by generating a stress (momentum-current density)
\begin{equation}
T^{ij}=\Lambda^{ijk\ell}u_{k\ell}+\eta^{ijk\ell}\dot{u}_{k\ell},\;\;\;\;\;\;\;\; u_{k\ell}=1/2(\partial_k u_\ell+\partial_\ell u_k)\label{eq:stress}
\end{equation}\noindent where repeated indices are always summed, $T^{ij}$ is the stress tensor (momentum current density), $\Lambda^{ijk\ell}$ is the elasticity tensor which relates stress to the strain $u_{k\ell}$ (\emph{i.e.} a generalization of Hooke's law) and $\eta^{ijk\ell}$ is the viscosity tensor relating stress to the strain rate/velocity gradient $\dot{u}_{k\ell}$ (\emph{i.e.} a velocity dependent frictional force). See Figure \ref{fig:elastic}a for an illustration of a lattice elastic medium and a displacement field.

A non-zero strain tensor indicates that the (spatial) geometry of the elastic medium has been distorted. The geometric characterization of the lattice is contained in the metric tensor which determines the distance between lattice points. In the ordered reference state shown in Figure \ref{fig:elastic}a the metric tensor is just $g_{ij}=\delta_{ij}$ which implies that distances between sites are calculated in the usual Euclidean way. When the material is strained, the spatial metric tensor is modified to become $g_{ij}=\delta_{ij}+2u_{ij}$\cite{landauElasticity}, which is what is meant when we say the geometry is deformed. Static lattice deformations affect the electronic behavior since the bonds are deformed. For electrons described by the Schrodinger equation at low-energy, the Hamiltonian is modified to become (to linear order in strain)
\begin{equation}
H=\frac{p^2}{2m}\to \frac{p_{i}g^{ij}(x)p_j}{2m}=\frac{p^2}{2m}-2u^{ij}(x)\frac{p^2}{2m}+i\hbar(\partial_{i}u_{ij}(x))\frac{p_j}{m}
\end{equation}\noindent where $g^{ij}(x)$ is the inverse of the metric tensor which depends on position via the contribution of the strain tensor. Thus, depending on the spatial profile of the strain, the electron spectrum can be drastically modified.

While the strain/metric based elasticity theory is quite successful, it is not general enough to model all of the electronic structure effects arising from the coupling of materials with spin-orbit coupling to geometric deformations. What is needed is a more fundamental field: the frame field  $\e_a$ in $d$ spatial dimensions where $a=1,2,\ldots d$ labels each vector of the frame (with components $\e_{a}^{i}$). The frame-field is a set of $d$ vectors residing on \emph{each} lattice site,  and heuristically  encodes the local bond stretching (through the vector lengths) \emph{and} the local orbital orientation (through their relative angles on each site). As we will see later, in many instances it is more natural to consider the co-frame field $e^a$ which is a local basis of $1$-forms that are dual to the vectors $\e_b$ (\emph{i.e.} they satisfy $e^a(\e_b)=\delta^{a}_{b}$).  For the reference state shown in Figure \ref{fig:elastic}a the reference frame fields are orthonormal vectors which are aligned with the crystal axes. The distances between lattice sites, \emph{i.e.} the (inverse) metric tensor is determined from the frame fields via $g^{ij}(x)=\delta^{ab}\e^{i}_{a}(x)\e^{j}_{b}(x)$\cite{MTWbook}. It is easy to see that if the frame fields are orthonormal at each site then $g^{ij}=\delta^{ij}$ as expected. The key relationship between the metric and the frame is that we can \emph{locally} rotate the frame at each site by any $SO(d)$ rotation matrix $R$ and we get the \emph{same} metric back: 
\begin{equation}
\tilde{g}^{ij}=\delta^{ab}(R_{a}^{c}(x)\e_{c}^{i}(x))(R_{b}^{d}(x)\e_{d}^{j}(x))=R_{a}^{c}(x)R_{a}^{d}(x)\e_{c}^{i}\e_{d}^{j}=\delta^{cd}\e_{c}^{i}\e_{d}^{j}=g^{ij}
\end{equation}\noindent since $RR^{T}=\mathbb{I}.$ This implies that an elasticity theory determined completely from the metric does not capture local orbital deformations since each different local orbital orientation yields the \emph{same} metric tensor. However, electrons with SOC propagating in a lattice will be sensitive to the local orbital orientation, which is exactly why a frame field must be introduced to couple these materials to geometric perturbations. This modification to elasticity theory is closely related to so-called micro-polar or `Cosserat' elasticity\cite{eringen1967,Hehl2007}.

At this point it is useful to explicitly show how the frame field enters spin-orbit coupled Hamiltonians. The low-energy description of two such systems are given by the Dirac Hamiltonian (which represents, for example, topological insulators) \cite{bernevig2006c,zhang2009} and the Luttinger Hamiltonian (which represents, for example, the upper-most valence bands of III-V semiconductors) \cite{winklerbook,murakami2003}:
\begin{eqnarray}
H_D&=&v\sum_{i,a} p_{i}\e^{i}_{a}\Gamma^{a}+m\Gamma^0\\
H_L&=&\delta^{ab}\frac{p_{i}\e^{i}_{a}\e^{j}_{b}p_j}{2m}+\alpha \left(p_{k}\e^{k}_{a}S^{a}\right)\left(p_{\ell}\e^{\ell}_{b}S^{b}\right)=\frac{p_i g^{ij}p_j}{2m}+\alpha (p_k\e^{k}_{a})(p_\ell \e^{\ell}_b)S^a S^b
\end{eqnarray}\noindent for Dirac matrices $\Gamma^{a}$, spin-3/2 matrices $S^a,$ and parameters $v,m,\alpha.$ Hence, the prescription is to replace terms of the form $p_iM^{i}$ for a matrix $M^i$, which arise naturally in materials with SOC, with $\sum_{a} p_i \e^{i}_{a}M^a.$ Note that for $H_L,$ since $S^a S^b\neq \delta^{ab},$ the quadratically dispersing Luttinger model is indeed affected by the local orbital orientation since it couples to more than just the metric tensor. The effects of the frame field are thus not limited to the linearly dispersing Dirac equation and affect any coupling between the direction of electron propagation $p_i$ and the spin/orbital degrees of freedom represented by $M^i.$ 

There are two complimentary interpretations of the (co-)frame-field which we will use. The first interpretation is in terms of familiar elasticity quantities, namely to first order in the displacement field, the co-frame and frame can be expanded as 
\begin{equation}
e^{a}_{i}=\delta^{a}_{i}+\frac{\partial u^{a}}{\partial x^i},\;\;\;\;   \e_{a}^{i}=\delta_{a}^{i}-\frac{\partial u_{a}}{\partial x_i}
\end{equation}\noindent where $\partial_{i} u^a\equiv w_{i}^{a}$ is the \emph{distortion tensor} which is familiar from elasticity theory\cite{landauElasticity}. 
The quantity $w^{a}_{i}$ is effectively the unsymmetrized strain tensor and contains information about local rotations through the anti-symmetric combination $M_{ij}=\delta_{ia}w^{a}_{j}-\delta_{ja}w^{b}_{i}.$  The distortion tensor also contains information about dislocations through the line-integral
\begin{equation}
 \oint_{C} w^{a}_{i}dx^i=\oint_{C} du^{a}=-b^a
\end{equation}\noindent where $b^a$ are the components of the total Burgers vector of the dislocation(s) enclosed within the curve $C$ (see Figure \ref{fig:elastic}b for an example)\cite{landauElasticity}. 

For point-like dislocations in 2d we can write $de^{a}=-b^a\delta^{(2)}(x)$ from Stokes' theorem where $de^{a}$ is the exterior derivative of the $1$-form $e^a.$ This formula suggests a second description of the $e^a$ as a set of $d$ vector potentials. As a comparison, we know that for electrons in an electromagnetic vector potential we use the minimal coupling replacement $p_i\to p_i +qA_i$ which \emph{shifts} the momentum in the Hamiltonian, and we have already mentioned that the proper replacement for the frame field is to \emph{scale} momentum
\begin{equation}
p_i \to p_{i}\e^{i}_{a} = p_{i}\delta^{i}_{a}-p_{i}w^{a}_{i}=p_a-p_{i}w^{a}_{i}.
\end{equation} Comparing to the electromagnetic case, this shows that each frame-vector yields a vector potential that minimally couples to electrons via momentum \emph{i.e.} the momentum components are the \emph{charges} of these gauge fields. With this interpretation, dislocations are just the magnetic fluxes of these vector potentials, and the translation effect of a dislocation is just the Aharonov-Bohm effect for the co-frame vector potentials. In general we can construct the \emph{torsion} tensor, which, in the absence of curvature can be chosen to take the simple form of a field strength tensor of the co-frame vector potentials  
\beq
{T_{ij}}^a=\partial_i e^{a}_{j}-\partial_{j}e^{a}_{i}
\eeq
This has an extra index $a$ compared to the electromagnetic version $F_{ij},$ which labels the particular vector potential/co-frame potential. This is how ``torsion" naturally enters the discussion, and as we can see, it is intimately connected to dislocation density. 

Along similar lines, we must also consider \emph{disclination} defects which represent sources of geometric curvature (see figure \ref{fig:elastic}c). These are naturally described by introducing the \emph{spin connection} (or simply \emph{connection}) \({\omega^a}_{b}={{\omega_{i}}^a}_bdx^i\), which is an anti-symmetric matrix of 1-forms. In analogy with the dislocation case, the connection contains information about the disclination (Frank) angle \({\theta^a}_{b}\) along a closed curve \(C\) through the line integral
\beq
{\theta^a}_{b} = -\oint_C {{\omega_{i}}^a}_bdx^i
\eeq
The connection is thus simply the matrix of \emph{non-Abelian} vector potentials which correspond to local rotations. The field strength for these vector potentials 
\beq
{R^a}_{b;ij} = \partial_i{{\omega_{j}}^a}_b-\partial_j{{\omega_{i}}^a}_b+{{\omega_{i}}^a}_c{{\omega_{j}}^c}_b-{{\omega_{j}}^a}_c{{\omega_{i}}^c}_b
\eeq
is called the \emph{curvature} 2-form. As we will see later, the connection couples to particles with non-trivial spin and leads to very important physical effects. We also mention that there exist other elastic defects like orbital-twisting defects that can be produced in a strain-free lattice with a trivial metric but non-trivial frame (\emph{e.g.} a torsional skyrmion\cite{randono2011}). 

\begin{figure}[t]
\begin{center}
\includegraphics[width=6.45in]{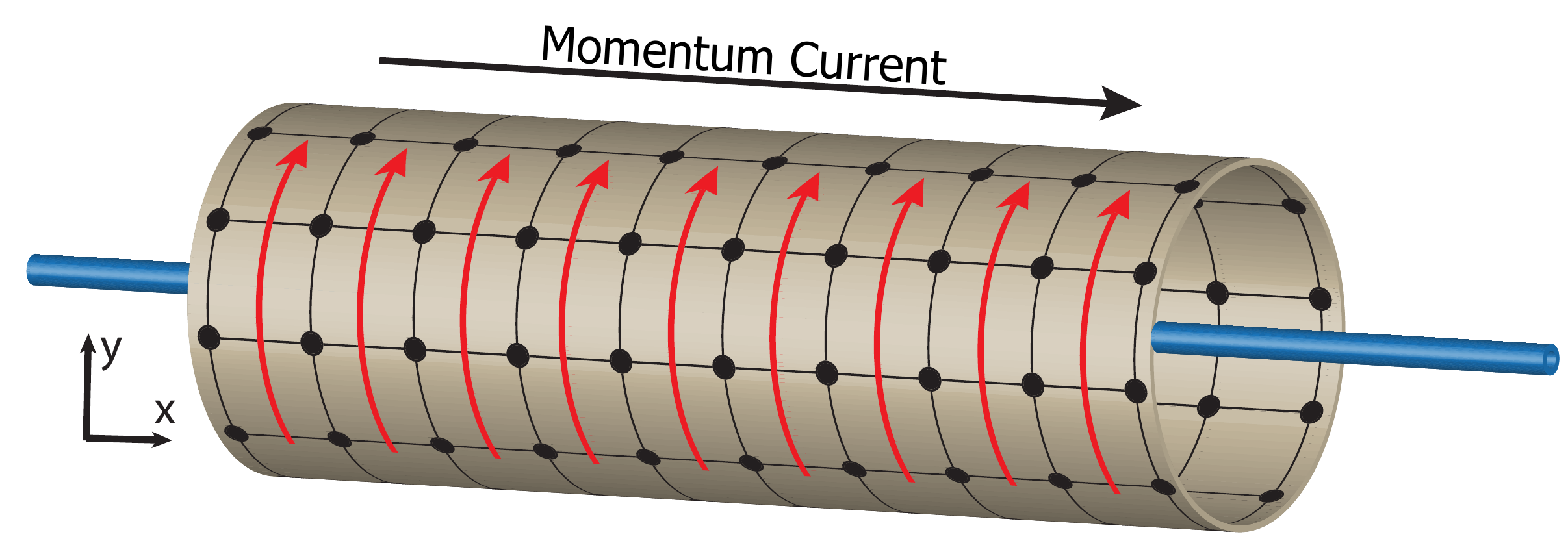}
\caption{Laughlin gauge argument for torsion: Thought experiment  with an insertion of torsion flux \emph{i.e.} a dislocation into cylindrical hole, equivalent to shrinking or enlarging the cylinder in the $y$-direction as a function of time. Non-zero dissipationless viscosity causes transfer of $p_y$-momentum in the $x$-direction, \emph{i.e.} a momentum current perpendicular to time-dependent strain.}\label{fig:twist}
\end{center}
\end{figure}

With the background theory now set up, we will move on to discuss the current state of the field of topological VE response, and some of the open questions which we are proposing to study. 
The first calculation of a topological VE response was the work of Avron \emph{et al.} which showed that a dissipationless viscosity is present in integer quantum Hall states\cite{avron1995,levay1997,avron1998}. The work was not followed up on until over a decade later when Read showed that fractional quantum Hall states (and chiral superfluids), being time-reversal breaking fluids,  also exhibit such a viscosity, and that the response is quantized if rotation symmetry were preserved\cite{read2009,read2011}. The quantization is not conventional because it involves a product of a quantized universal factor and a non-universal inverse area scale, which in the quantum Hall effect is related to the size of area quantization, \emph{i.e.} the magnetic length ($\ell_{B}^2$). So it is only the ratio of the viscosity to the density which is quantized with rotational symmetry.  The notion of a quantized viscosity, which is relevant to our current work, is less clear in the case of topological insulators, and the issue has not been settled (even in continuum, rotationally invariant models)\cite{hughes2011}. Soon after, Haldane showed that rotation symmetry is not a necessary ingredient for defining a quasi-universal property associated with quantum Hall states. Namely he showed that the viscosity is related to a universal property of an \emph{unreconstructed} quantum Hall edge: the edge dipole moment\cite{haldane2009A}. For these systems the viscosity, denoted $\zeta_H,$ is a quantized multiple of $\hbar/\ell_{B}^{2}$ where $\ell_B$ is the magnetic length. This quantity has units of angular momentum density, or momentum per unit length, or dynamic viscosity (force/velocity), and interestingly, it depends on a non-universal length scale which varies when the magnetic field is tuned. In fact, one even can remain on the same Hall plateau with fixed Hall conductance, and tune the field so that the viscosity changes. When rotation symmetry is present, conserved angular momentum can be transferred between edges via an applied torque (\emph{e.g.} due to the electric field generated from perpendicular applied flux). The amount of transferred angular momentum does not depend on $\ell_B,$ and is given by the quantized multiple of $\hbar$ appearing in $\zeta_H.$ The same is true of the edge dipole moment, which is also independent of $\ell_B$ for unreconstructed edges, and is the same universal number multiplying $\hbar.$

This quantization emerges quite naturally in the Landau level problem where the quantum Hall effect is generated by an external magnetic field. However, the situation is more subtle and complicated when the quantum Hall effect is generated by a topological band structure which can naturally furnish multiple length scales. We will focus on this type of system to study the impact that a combination of geometry and topology will have in band theory. The topological viscosity has been calculated in a (properly regularized) continuum model for the Chern insulator, \emph{i.e.} the massive Dirac Hamiltonian in 2+1d \cite{hughes2011}; the details of this calculation will be presented in section \ref{section 5}. In flat space the regularized value was found to be $\zeta_H=\frac{\hbar}{8\pi\xi^2}$ where $\xi=\hbar v/2m$ is the length scale induced by the Dirac mass $m$ (with units of energy) for a material with a Fermi-velocity (speed of light) $v.$ In spaces with constant Riemann curvature one finds a universal topological correction which yields a viscosity  \begin{equation}\zeta_H=\frac{\hbar}{8\pi\xi^2}-\frac{\hbar}{12A}\chi_{\Sigma}\end{equation}\noindent where $A$ is the spatial area of the system and $\chi_{\Sigma}$ is the Euler characteristic.  In relation to the discussion of elasticity theory above, the non-zero viscosity coefficient produces a Chern-Simons response for the \emph{co-frame fields}:
\begin{equation}
S_{eff}[e^{a}]=\frac{\zeta_H}{2}\int d^2 x dt\ \epsilon^{\mu\nu\rho}e_{\mu}^{a}\partial_{\nu}e_{\rho}^{b}\eta_{ab} \label{hallviscositytake1}
\end{equation}
\noindent where $a,b=0,1,2,$ and $\eta_{ab}={\textrm{diag}}[-1, 1, 1]$ is the flat-space Minkowski metric. This is essentially multiple copies of the conventional Abelian Chern-Simons term, one for each of the co-frame fields (including the co-frame in the time direction). As shown in Ref. \cite{hughes2011}, if we calculate the electronic contribution to the \emph{stress current}  
\beq
(J^{a})^{\mu}=\frac{1}{\mathrm{det}(e)}\frac{\delta S_{eff}}{\delta e_{a;\mu}}=\frac{\zeta_H}{2}\epsilon^{\mu\nu\rho}{T^a}_{\nu\rho}
\eeq
one finds that electron \emph{momentum-density} is bound at dislocation defects and \emph{momentum-current} is generated perpendicular to any velocity-gradients/strain-rates (see Figure \ref{fig:twist}a for a picture of the latter). This is completely analogous to the \emph{charge density} bound to magnetic flux and \emph{charge current} produced by electric fields (or time-dependent fluxes) in the quantum Hall Chern-Simons response. Additional discussions of the Hall viscosity in topological insulators and the relation to lattice deformations and the electron-phonon coupling\cite{chung2012}, a spin-Hall viscosity\cite{kimura2010}, and a Streda-like formula for the viscosity\cite{kimura2012} have been carried out. As the reader might have guessed, there is also a Chern-Simons response for the \emph{spin connection}, which schematically is of the form\footnote{We will see later in this paper that the effective spin-connection in the Dirac model gets repackaged.}
\begin{equation}
S_{eff}[\omega_{ab}]=\frac{\kappa_H}{2}\int d^2 x dt\ \epsilon^{\mu\nu\rho}\left({{\omega_{\mu}}^a}_b\partial_{\nu}{{\omega_{\rho}}^b}_a+\frac{2}{3}{{\omega_{\mu}}^a}_b{{\omega_{\nu}}^b}_c{{\omega_{\rho}}^c}_a\right) \label{thermalhalltake1}
\end{equation}
where the coefficient \(\kappa_H\) for the Chern insulator will be computed in section \ref{sec: chiral gravity}. This gives rise to a \emph{spin current}\footnote{In the fully Lorentz covariant description that we will present, there is another term in the spin current proportional to \(\zeta_H\).}
\beq
({J^a}_b)^{\mu}=\frac{1}{\mathrm{det}(e)}\frac{\delta S_{eff}}{\delta {{\omega_{\mu}}_a}^b}=\frac{\kappa_H}{2}\epsilon^{\mu\nu\rho}{R^a}_{b;\nu\rho}
\eeq
and we thus find electron \emph{spin-density} bound to disclination defects. In the case of vanishing torsion one can determine the spin connection from the frame-field and one subsequently finds that the spin-connection Chern-Simons term leads to thermal currents in response to gravitational tidal forces\cite{stoneGCS}

The principal issue we deal with in this work is developing a physical understanding of the viscosity, and in general, the gravitational response theory. In flat space the viscosity $\zeta_H$ does not appear to be quantized, or even universal, which is \emph{very} strange in light of all the previous results on topological responses in topological insulators, and thus requires explanation\cite{thouless1982,qi2008b}. In this paper we approach these issues by considering the bulk-boundary correspondence between momentum-transport in the bulk and anomalous currents in the chiral edge states. We develop a picture analogous to the charge response for the quantum Hall effect in terms of bulk Chern-Simons response and a boundary chiral-anomaly. To address these issues we must develop more precise language than we have used in this more informal section; we do this now. In particular, we will treat the \emph{co-frame} and the \emph{connection} on the same footing, and more importantly restore Lorentz covariance. As a note, a condensed-matter minded reader might first read sections \ref{section 5}, \ref{sec: chiral gravity}, and \ref{sec:spectralflow} to get some picture of the physics before tackling the more technical, but essential, discussions of the other sections.



\section{Geometry with Torsion} \label{section 3}
\newcommand{\LCconn}{{\omega^{\kern-6pt{\tiny\circ}\kern2pt}}}
\newcommand{\LCTor}{{T^{{}^{\kern-6pt{\tiny\circ}\kern2pt}}}}
\newcommand{\LCcurv}{{R^{{}^{\kern-5.5pt{\tiny\circ}\kern2pt}}}}
\newcommand{\LCcov}{{D^{{}^{\kern-6pt{\tiny\circ}\kern2pt}}}}

Gravity is usually described as a theory of metrics, corresponding to a measure of invariant distance
\beq
ds^2=g_{\mu\nu}dx^\mu dx^\nu
\eeq
where $x^\mu$ are local coordinates on a manifold. 
We can package the information contained in the metric (and more in fact) into the components of a co-frame, a \textsl{local} basis of 1-forms $e^a=e^a_\mu dx^\mu$ on the manifold. Equivalently, we can regard $e^a$ as a local section of the oriented co-tangent bundle of the manifold. The metric is related to the components of the 1-forms via 
\beq
e^a_\mu e^b_\nu\eta_{ab}=g_{\mu\nu}
\eeq
where $\eta_{ab}$ are the components of the Lorentz-invariant Minkowski metric. We will denote the dual set of \emph{frame vector fields} as $\e_a$, with $e^a(\e_b)=\delta^a_b$. To translate (co-)tangent bundle data from point to point on the manifold, we need a connection or covariant derivative $\nabla$. Conventionally we write the translation of the frame along a vector field $\underline{X}$ as
\beq
\nabla_{\underline{X}} e^a=- {\omega^a}_b(\underline{X}) e^b\label{introspinconn}
\eeq
where we have introduced the components of the spin connection ${\omega^a}_b$, which we regard as a set of 1-forms. In a basis of local coordinates this equation can be written as $X^{\mu}\nabla_{\mu}e_{\nu}^{a}=-X^{\mu}{{\omega_{\mu}}^{a}}_{b}e^{b}_{\nu}.$ The spin-connection can be thought of as a non-Abelian gauge field that couples to the rotation and Lorentz transformation generators. Throughout our work, we will make one assumption about this connection, which is that it is {\it metric compatible}.  In metric terms, this means that the metric is covariantly constant $\nabla_X g=0$, but using the relationship between the metric and the co-frame and the definition (\ref{introspinconn}), it also corresponds to the spin connection being valued in the orthogonal group,\footnote{We will work in $d=D+1$ spacetime dimensions, so the relevant orthogonal group is $SO(1,D)$.} {\it i.e.}, $\omega_{ab}=-\omega_{ba}$ (where $\omega_{ab}\equiv \eta_{ac}{\omega^c}_b$). Under a local change of basis ({\it i.e.}, a local Lorentz transformation) $e^a\mapsto {\Lambda^a}_be^b$, the connection transforms as\footnote{Note that we are reserving the term `Lorentz transformation' for these local changes of basis for the orthonormal co-frame. These should not be confused with (linear) diffeomorphisms, which are local changes of the coordinates.}
\beqn
{\omega^a}_b &\mapsto& {(\Lambda \omega\Lambda^{-1}-d\Lambda\Lambda^{-1})^a}_b
\eeqn
Thus ${\omega^a}_b$ is the `gauge field' for local Lorentz transformations.
The curvature 2-form, or field strength, of the connection
\beqn
{R^a}_b&=&d{\omega^a}_b+{\omega^a}_c\wedge {\omega^c}_b.
\eeqn
transforms linearly
\beq
{R^a}_b\mapsto {(\Lambda R\Lambda^{-1})^a}_b.
\eeq
The components of the curvature 2-form give the Riemann tensor, $R_{cd}=\frac12 R_{ab;cd} e^a\wedge e^b$.
If we denote the covariant derivative acting on (local) Lorentz tensors by $D$, the torsion 2-form is defined as
\beqn
T^a&=& De^a\equiv de^a+{\omega^a}_b\wedge e^b
\eeqn
Torsion also transforms linearly under local Lorentz transformations.\footnote{The Bianchi identities are
\beqn
{DR^a}_b&\equiv& {dR^a}_b- {R^a}_c\wedge{\omega^c}_b+{\omega^a}_c\wedge {R^c}_b=0\\
DT^a&\equiv &dT^a+{\omega^a}_d\wedge T^d={R^a}_d\wedge e^d
\eeqn
If the torsion vanishes, the latter corresponds to a symmetry property of the Riemann tensor.
}
\beq
T^a\mapsto {\Lambda^a}_b T^b.
\eeq
We write the components of the torsion 2-form as $T^c=\frac12 {T^c}_{ab} e^a\wedge e^b$.

A very basic property of the connection, is that it satisfies the following translation algebra
\beqn
\left[\nabla_a,\nabla_b\right] &=& -T^c_{ab}\nabla_c+R_{cd;ab}J^{cd}\label{eq:transalg}
\eeqn
where $J^{cd}$ is the generator of rotations. We will see an explicit representation of this algebra later in the paper. The left hand side can be interpreted as successive translations along $\e_b, \e_a, -\e_b, -\e_a$, and thus we see that the components of the torsion tensor correspond to the non-closure of these successive translations by an extra translation, while the components of the Riemann tensor imply that a rotation is also involved. 

In classical general relativity (GR), a basic property of the theory is that the torsion is taken to vanish; this is one manifestation of the equivalence principle.  In fact, there is a \emph{unique} connection, the Levi-Civita connection ${\LCconn^a}_b$, with this property which is determined entirely by the co-frame alone ({\it i.e.}, the metric). Indeed in the familiar Einstein-Hilbert Lagrangian formulation of GR, the torsion vanishes as a constraint. In other formulations (the first-order or Riemann-Cartan formulations), $e^a$ and ${\omega^a}_b$ are regarded as independent degrees of freedom and the torsion may then vanish by equations of motion (for suitable choice of matter field configurations). In the latter formalism, one can envisage including sources that would induce torsion, much as the usual sources induce curvature. It should be emphasized though that in our context, we regard $e^a$ and ${\omega^a}_b$ as background fields, with no dynamics of their own.

Given the form of the translation algebra (\ref{eq:transalg}),  the vanishing of torsion in fact corresponds to a {\it choice of state}. As in the previous section we can consider an elastic medium given by a (space-time) lattice $\Lambda$. We will typically be interested in continuum limits, giving rise to a continuum quantum field theory, in the presence of a variety of background fields (so that we can study various transport properties). At each point in the lattice, we have defined a frame, whose magnitudes are tied to the (local) lattice spacing. The commutator of translations on the lattice is defined by hopping along a square path; failure to return to the starting position corresponds to the path encircling a {\it dislocation} of the lattice, and the magnitude and direction of the translation determines the {\it Burgers' vector} $\underline{b}$ of the dislocation.  There exist two primary types of dislocations: (i) an edge dislocation with $\underline{b}$ perpendicular to the tangent vector of the dislocation line (b) a screw dislocation with $\underline{b}$ parallel to the dislocation line (only exists in 3+1-d or higher). An example of the former is shown in Fig. \ref{fig:elastic}b.
Now consider a continuum limit. If the limit is taken in such a way that a {\it density of dislocations} $\underline{b}(x)$ is obtained, we should associate this with non-zero torsion in the continuum theory. Lattice dislocations correspond to point sources of torsion. The frame is rotated if the path encircles a {\it disclination} and continuum limits yielding a density of disclinations corresponds to curvature. Disclinations are significant if and only if the field in question carries a non-trivial Lorentz representation (that is the generator $J^{ab}$ is non-zero), {\it i.e.}, it carries spin. The effects of dislocations do not carry this requirement.

Thus, in condensed matter systems coupled to elastic media, we conclude that the presence of curvature and torsion in the continuum limit corresponds to a choice of state. Since both curvature and torsion are present, the nature of the background is determined not just by the metric, but by both the co-frame and connection. As we will show in detail below, this corresponds to the presence of independent  Lorentz and diffeomorphism currents (whereas in the absence of torsion, these reduce to just the conventional stress-energy tensor). 
However, even in the absence of torsion in the ground state of the system, torsional perturbations should be also considered in the context of transport properties. Studying effective actions\footnote{We use the term `effective action' here interchangeably with `generating functional'. The latter term is most appropriate, as indeed, the use of the effective action is that it encodes the correlation functions of currents.} of a given field theory in the presence of background co-frame and connections is equivalent to studying the correlation functions of the these currents, as the backgrounds correspond to sources for the current operators. In some cases, torsion appears in terms in this effective action, with coefficients that are physically meaningful. For example,  in 2+1 dimensions in the presence of time-reversal symmetry breaking, there is a non-dissipative transport coefficient (the Hall viscosity) that is the coefficient of a Chern-Simons-like term involving torsion, as mentioned in the previous section\cite{hughes2011}.

It is convenient to introduce some additional notation. As indicated above, given a co-frame $e^a$, there is a uniquely determined {\it Levi-Civita connection} ${\LCconn^a}_b$ whose torsion vanishes. 
We define the {\it contorsion} ${\con^a}_b$ via
\beq
{\omega^a}_b={\LCconn^a}_b+{\con^a}_b
\eeq
so\footnote{We define $\LCcov$ as the LC covariant derivative, ${(\LCcov C)^a}_b=d{C^a}_b+{\LCconn^a}_c\wedge {C^c}_b+{C^a}_c\wedge {\LCconn^c}_b$ and $\LCcurv$ as the LC curvature ${\LCcurv^a}_b=d{\LCconn^a}_b+{\LCconn^a}_c\wedge {\LCconn^c}_b$.}
\beqn
T^a&=&{\con^a}_b\wedge e^b\\
{R^a}_b&=&{\LCcurv^a}_b+{(\LCcov \con)^a}_b+{\con^a}_c\wedge {\con^c}_b
\eeqn
Note also that the contorsion is a Lorentz tensor. Generally, we will regard $e^a$ and ${\omega^a}_b$ as independent. For later use, we will also define
\beq
H=\frac{1}{3!}H_{abc}e^a\wedge e^b\wedge e^c = e^a\wedge T^b\eta_{ab}
\eeq
where $H_{abc}=-3! C_{[a;bc]}=3T_{[bc;a]}$. $H$ is not in general a closed form, and hence we define the Nieh-Yan 4-form
\beq
N=dH=T^a\wedge T^b\eta_{ab}-R_{ab}\wedge e^a\wedge e^b.
\eeq

\section{Generic Properties and Symmetries of Fermions Coupled to Torsion} \label{section 4}

In this section, we will discuss various aspects of fermions on a generic background, mainly focussing on the role played by torsion. We are studying Dirac models since they represent the minimal continuum models of topological insulators in any dimension. In the following, we assume that we are in a $d=(D+1)$-dimensional space-time with co-frame $e^a$ and connection ${\omega^a}_b$ with a mostly-plus metric. Results and conventions for spinors and Clifford (Dirac) algebra can be found in the appendix. 

\subsection{Dirac fermions}

The Dirac action may be written as\footnote{The (Lorentz and gauge) covariant derivative of the Dirac spinor is $\nabla\psi=d\psi+\frac14 \omega_{ab}\gamma^{ab}\psi+A\psi$, where $A$ is an appropriate (non-Abelian) gauge connection. We note also that the invariant form of the action, eq. (\ref{DiracActionInv}), does not involve the frame $\underline{e}_a$ dual to $e^a$. }
\beqn
S[\psi; e,\omega]
&=&
\frac{1}{D!}\int \epsilon_{a_1\ldots a_d}e^{a_1}\wedge \ldots\wedge e^{a_{D}}\wedge \left[ \frac{1}{2}\overline\psi \gamma^{a_d}\nabla\psi-\frac{1}{2}\overline{\nabla\psi}\gamma^{a_d}\psi- e^{a_d}\overline\psi m\psi\right]
\label{DiracActionInv}\\
&=&\int d^dx\det e \left[ \frac{1}{2}\overline\psi \gamma^a\nabla_{\underline{e}_a}\psi-\frac{1}{2}\overline{\nabla_{\underline{e}_a}\psi}\gamma^a\psi-\overline\psi m\psi
\right] \label{DiracAction}
\eeqn
We have written the action in this way as it is precisely real (written in other ways, the action might be real up to the addition of a boundary term). In odd space-time dimensions, $m$ is real, and its sign will play a central role in determining the character of the resulting insulating state. In even space-time dimensions $m$ is essentially complex if no additional discrete symmetries are imposed ($m\to me^{i\theta \gamma_5}$, where $\gamma_5$ is the chirality operator). In addition, when torsion is non-zero, there is an additional term\footnote{There are actually two other terms at the same level of power counting. The first, of the form $i\int\det e \left[T^a(\underline{e}_b,\underline{e}_c)\overline\psi [\gamma_a,\gamma^{bc}]\psi-2\nabla_{\underline{e}_a}(\overline{\psi}\gamma^a\psi)\right]$ is Nieh-Yan-Weyl invariant (see below), but a total derivative. The second, of the form $i\int\det e\ T^a(\underline{e}_b,\underline{e}_c)\overline\psi [\gamma_a,\gamma^{bc}]\psi$, is redundant (it can be absorbed into the definition of a $U(1)$ gauge field).
} that can be added to the action, of the form
\beq
S_T[e,\omega]=\frac{1}{16}\alpha\int \det e\  T^a(\underline{e}_b,\underline{e}_c)\overline\psi \{\gamma_a,\gamma^{bc}\}\psi.
\eeq

The classical equation of motion for the spinor field involves the Dirac operator
\beq\label{DiracOp}
\cDsl=\gamma^ae_a^\mu\left(\pa_\mu+A_\mu^At_A+\frac14 \omega_{\mu;bc}\gamma^{bc}+B_\mu\right)+\frac18\alpha T_{bc;a}\gamma^{abc}
\eeq
where $B_a\equiv \frac{1}{2}T^b(\underline{e}_a,\underline{e}_b)=-\frac12{{\con}^b}_a(\underline{e}_b)$. 
The $B$ term arises upon integration by parts in deriving the equations of motion. We have included here for completeness a non-Abelian gauge field (if the spinor is in a gauge representation $t_A$) and we note that the torsional $B$-term enters in such a way that it looks like it corresponds to an additional gauge field. It is not of course independent of the spin connection, but does vanish with the torsion. In fact, as explained in \cite{NY}, the classical theory possesses a corresponding background scaling symmetry when $m=0$ under which the fields and background transform as
\beqn
&&e^a(x)\mapsto e^{\Lambda(x)}e^a(x),\ \ \ \ {\omega^a}_b(x)\mapsto {\omega^a}_b(x),\label{NYC1}\\
&& \psi(x)\mapsto e^{-(d-1)\Lambda(x)/2}\psi(x),\ \ \ \ \ \cDsl\mapsto e^{-\Lambda} (e^{-(d-1)\Lambda/2}\cDsl e^{(d-1)\Lambda/2}).\label{NYC2}
\eeqn
We note that this implies 
\beq
T^a\mapsto e^\Lambda\left(T^a+d\Lambda\wedge e^a\right)
\eeq
and hence
\beq
B_a=\frac{1}{2}T^b(\underline{e}_a,\underline{e}_b)\mapsto e^{-\Lambda}\left( B_a+\frac{d-1}{2}\underline{e}_a(\Lambda)\right)
\eeq
If we introduce a 1-form $B\equiv B_a e^a$, then this is equivalent to\footnote{Note that the invariance of ${\omega^a}_b$ implies that the contorsion transforms as
\beq
{C^a}_b\mapsto {C^a}_b-d\Lambda(\underline{e}_b)e^a+\eta^{ac}\eta_{bd}d\Lambda(\underline{e}_c)e^d
\eeq
and the LC connection transforms oppositely.
}
\beq
B\mapsto B+\frac{d-1}{2}d\Lambda
\eeq
which is the transformation of an Abelian ($\mathbb{R^+}$, not $U(1)$) connection.

We will refer to this as the Nieh-Yan-Weyl (NYW) symmetry.
Note that this is {\it not} the Weyl symmetry of the metric theory, because in that case, $\omega$ must transform in order that the torsion remain zero. In our case, the Weyl symmetry (at least as far as the {\it Dirac operator} is concerned) corresponds to a complexification of a $U(1)$ symmetry. In addition, the classical Dirac theory also has the usual background diffeomorphism, local Lorentz, and gauge symmetries, which we will discuss below. 

Another way to write the Dirac operator is in terms of the Levi-Civita connection, and the totally antisymmetric part of the contorsion
 \beqn\label{DiracOpH}
\cDsl&=&\gamma^ae_a^\mu\left(\pa_\mu+A_\mu^At_A+\frac14 \LCconn_{\mu;bc}\gamma^{bc}\right)+\frac14 \con_{a;bc}\gamma^a\gamma^{bc}+B_a \gamma^a+\frac18\alpha T_{bc;a}\gamma^{abc}\\
&=&\gamma^ae_a^\mu\left(\pa_\mu+A_\mu^At_A+\frac14 \LCconn_{\mu;bc}\gamma^{bc}\right)-\frac{1-\alpha}{4}\frac{1}{3!}H_{abc}\gamma^{abc}
\eeqn
where we have done some $\gamma$-matrix algebra (see Appendix) and defined $H_{abc}=-3! C_{[a;bc]}$. We will alternately regard $H_{abc}$ as the components of a 3-form $H=\frac{1}{3!}H_{abc}e^a\wedge e^b\wedge e^c$ or as the components of a vector-valued 1-form, $H_{ab}=H_{abc}e^c$. We note that the parameter $1-\alpha$ determines the coupling of (the antisymmetric part of) torsion to the fermions. Thus we can regard it as `torsional charge', and write $ 1-\alpha=\torcplg$. For convenience, we will set $\torcplg = 1$ throughout most of the paper, except in sections \ref{subsec: hall viscosity} and \ref{sec: chiral gravity}, where it is illuminating to resurrect it.

Since the Dirac theory is quadratic in fermion fields, the partition function in the quantum theory is obtained by performing a path integral over fermions 
\beq
Z(A,e^a,{\omega^a}_b;m)=\det (\cDsl-m)
\eeq
The diffeomorphism, local Lorentz, and gauge symmetries of the Dirac theory remain unaffected by perturbative (\emph{i.e.} local) anomalies upon quantization in arbitrary dimension. In odd dimensions the NYW symmetry at \(m=0\) is also non-anomalous. At \(m\neq 0\), the NYW symmetry is explicitly broken.
Additionally, the mass term also breaks parity invariance. In this paper we will mainly be interested in the quantum effective action for 2+1 dimensional Dirac fermions $S_{eff}[e,\omega,A]=-\mathrm{ln}\;\det (\cDsl-m)$. We denote the parity-violating piece of the effective action as  $S_{odd}[e,\omega,A].$ In the absence of torsion, symmetry considerations severely constrain the form of parity odd terms. For example in $d=3$, we have the Chern Simons terms
\beq
S_{odd}[e,\lcw,A]=\frac12\int\left(\sigma_H\;A\wedge dA + \kappa_H\mathrm{tr}\;(\LCconn\wedge d\LCconn+\frac{2}{3}\LCconn\wedge \LCconn\wedge \LCconn)\right) \label{CStorsionfree}
\eeq
The coefficient \(\sigma_H\) is called the \textsl{Hall conductance}, while \(\kappa_H\), the coefficient of the gravitational-Chern-Simons term, is related to the 2+1 dimensional Immirzi parameter. Non-zero torsion allows us to construct additional terms like 
\beq
\frac{1}{2}\int \zeta_H\; e^a\wedge T_a \label{CStorsional}
\eeq
This term was discussed in a slightly different guise in section \ref{section 2} (see Eq \ref{hallviscositytake1}); the coefficient \(\zeta_H\) is called the \textsl{Hall viscosity}. 
Additionally, the effective action also has parity even terms of the form 
\beqn
S_{even}[e,\omega,A] &=& \frac{1}{2\kappa_N}\int \left(\epsilon_{abc}e^a\wedge \lcR^{bc}-\frac{3\gamma^2}{2}H\wedge *H-\frac{\Lambda}{3}\epsilon_{abc}e^a\wedge e^b\wedge e^c\right)\nonumber\\
&=&\frac{1}{2\kappa_N}\int d^3x\; \mathrm{det}(e)\left(\lcR-\frac{\gamma^2}{4}H_{abc}H^{abc}-2\Lambda\right) \label{EinsteinHilbert2}
\eeqn
where \(\frac{\kappa_N}{8\pi}\) is the \textsl{Newton's constant}, \(\Lambda\) is the \textsl{cosmological constant}, and \(\gamma\) is a dimensionless parameter. We will examine $S_{eff}$ for the 2+1 Dirac model more closely in section \ref{sec: chiral gravity}. 


In even dimensions, it is also possible to couple chiral fermions to the frame and connection. The action is a straightforward modification of (\ref{DiracActionInv}, \ref{DiracAction})
\beqn
S_{\pm}[\psi; e,\omega]&=&\frac{1}{D!}\int \epsilon_{a_1\ldots a_d}e^{a_1}\wedge \ldots\wedge e^{a_{D}}\wedge \left[ \frac{1}{2}\overline\psi \gamma^{a_d}\nabla P_{\pm}\psi-\frac{1}{2}\overline{\nabla\psi}\gamma^{a_d}P_{\pm}\psi\right]
\label{ChiralActionInv}\\
&=&\int d^dx\det e \left[ \frac{1}{2}\overline\psi \gamma^a\nabla_{\underline{e}_a}P_{\pm}\psi-\frac{1}{2}\overline{\nabla_{\underline{e}_a}\psi}\gamma^aP_{\pm}\psi
\right] \label{ChiralAction}
\eeqn
with $P_{\pm} = \frac{1\pm\gamma^5}{2}$ being the chirality projection operators. The chiral theory also has the symmetries of the Dirac theory. However, all the symmetries are spoilt by perturbative anomalies upon quantization on generic backgrounds. Later in this paper, we will explore such chiral anomalous conservation laws for Lorentz, diffeomorphism, and gauge currents, particularly in $1+1$ dimensions, and their connection with $S_{odd}$ for the 2+1 Dirac model. We will see that while torsional terms like \eqref{CStorsional} leave \textsl{consistent} anomalies unaffected, they do modify the \textsl{covariant} anomalies. 
\subsection{Classical Ward identities}
In this section, we state the classical conservation laws for fermions coupled to the coframe, connection and a $U(1)$ gauge field.\footnote{In the rest of the paper, we will restrict the gauge group to $U(1)$ in favor of somewhat simpler notation. We will use the symbol \(q\) for the \(U(1)\) charge.} Although we will discuss these in the context of Dirac fermions (for arbitrary $d$), the results generalize in a straightforward manner to chiral fermions in even dimensions. Let us begin by defining the following currents 
\beqn
(J^{\mu}) &=& q\;\overline\psi \gamma^a\underline{e}^{\mu}_a\psi\\
(J_\mu)^a
&=& \frac12(\overline\psi\gamma^a \nabla_\mu \psi-\overline{\nabla_{\mu}\psi}\gamma^{a}\psi)
\\
{(J^\mu)^a}_b
&=&\frac14 \underline{e}^\mu_c\overline\psi {\gamma^{ca}}_b\psi \label{spin current}
\eeqn
which we will refer to as the \emph{charge} current, \emph{stress} current and \emph{spin} current respectively. These couple respectively to the \(U(1)\) gauge field, coframe, and spin connection in the classical action. In the absence of torsion the last two currents are not independent. The components of the current $J^a$ give the usual notion of the stress-energy tensor via
\beq
T_{\mu\nu}= J^a_\mu e^b_\nu \eta_{ab}
\eeq
Also note that the spin current \(J^{ab}_{\mu}\) vanishes in \(d=2\). It will be convenient to introduce the corresponding 1-forms $J = J_{\mu}dx^{\mu}$, $J^a = J^a_{\mu}dx^{\mu}$ and $J^{ab} = J^{ab}_{\mu}dx^{\mu}$. Invariance under \(U(1)\) gauge transformations implies that \(J\) is conserved, \emph{i.e.} $d*J=0$, which in components is the usual $\pa_\mu(\mathrm{det}(e) J^\mu)=0$.


\subsubsection{Diffeomorphisms}

The invariance of the classical action under local background diffeomorphisms follows immediately from writing it as the integral of a top form, as in \eqref{DiracActionInv}. We will take the action of local diffeomorphisms on fermions and background fields as 
\beq
\delta \psi = i_{\xi}\nabla\psi,\;\; \delta e^a = D\xi^a+i_{\xi}T^a,\;\; \delta \omega_{ab} = i_{\xi}R_{ab},\;\; \delta A = i_{\xi}F \label{covdiff}
\eeq
where $\xi$ is a vector field with compact support and $ i_{\xi}$ is the interior product of $\xi$ with a differential form. These transformations differ from ordinary diffeomorphisms by local gauge transformations, so we will refer to these as \textsl{covariant diffeomorphisms}. Using equations of motion for the fermions, the variation in the action under \eqref{covdiff} is given by 
\beq
\delta_{Diff.}S = \int \left[i_{\xi}F\wedge *J+(D\xi_a+i_{\xi}T_a)\wedge *J^a +i_{\xi}R_{ab}\wedge *J^{ab}\right]
\eeq
and so invariance of the action implies the classical Ward identity 
\beq 
D*J^a - i_{\underline{e}^a}T_b\wedge *J^b-i_{\underline{e}^a}R_{bc}\wedge *J^{bc}-i_{\underline{e}^a}F\wedge *J = 0 \label{ClassicalDiffWI}
\eeq 




\subsubsection{Local Lorentz transformations}

The spinors and background fields transform under an infinitesimal Lorentz transformation as
\beq
\delta\psi = \frac14\theta_{ab}\gamma^{ab}\psi,\;\delta e^a = -{\theta^a}_be^b,\;\delta{\omega^a}_{b} = -{(D\theta)^a}_b\label{loclor}
\eeq
Under \eqref{loclor}, the action changes by
\beqn
\delta_{Lor.} S 
=-\int \left[ D\theta_{ab}\wedge *J^{ab}+\theta_{ab} e^a\wedge *J^b\right]
\eeqn
The Ward identity is
\beq
D*J^{ab}-e^{[a}\wedge *J^{b]} = 0 \label{ClassicalLorWI}
\eeq



\subsubsection{Nieh-Yan-Weyl transformations}

The action on fermions and background fields is given by 
\beq
\delta\psi = -\frac{d-1}{2}\Lambda\psi,\;\delta e^a = \Lambda e^a,\;\delta \omega_{ab}=0
\eeq
Under $\delta \psi = -\frac{d-1}{2}\Lambda\psi$, the action transforms as
\beqn
\delta S_{NYW}
&=&-(d-1)\int \Lambda \left[\eta_{ab}e^a\wedge *J^b-m\ vol\ \overline\psi\psi\right]
\eeqn
The second term, where $vol$ is the volume form,  is present because the mass term explicitly violates the NYW symmetry. For \(m=0\), we have the Ward identity
\beq
\eta_{ab}e^a\wedge *J^b =0 \label{ClassicalNYWI}
\eeq
In components, this is ${T^{\mu}}_{\mu}=g^{\mu\nu}e^a_\mu J_\nu^b\eta_{ab}$, the trace of the stress-energy tensor. Thus in this sense, the NYW symmetry gives rise to the same conservation law as does Weyl invariance of the second-order formalism.

\subsection{Lichnerowicz-Weitzenbock Formula}
The heat kernel for the operator  $\cDsl^2$ will play a central role in some of our computations, so before moving on, let us briefly describe this operator. We begin by noting
\beqn
\cDsl^2&=&
\gamma^a(\nabla_a+B_a)\gamma^b(\nabla_b+B_b)\\
&=&\gamma^a\gamma^b(D_a+B_a)(D_b+B_b)\\
&=&\gamma^a\gamma^b {\cal D}_a{\cal D}_b
\eeqn
where $D_a$ is fully (Lorentz) covariant and ${\cal D}_a= D_{\underline{e}_a}+B_a$. In manipulating this expression we need various facts about the Clifford algebra (see Appendix) and we also encounter the commutators
\beqn
\left[D_a,D_b\right] &=& -T^c_{ab}D_c+\frac14 R_{cd;ab}\gamma^{cd}+iqF_{ab}\label{eq:transalgF}\\
\left[D_{[a},B_{b]}\right] &=& -\frac12T^c_{ab}B_c+\frac12 G_{ab}
\eeqn
where $G_{\mu\nu}=\pa_\mu B_\nu-\pa_\nu B_\mu$. 
Consequently, the Lichnerowicz-Weitzenbock formula takes the general form
\beqn
\cDsl^2&=&
\eta^{ab}{\cal D}_a{\cal D}_b-\frac14 R+\frac{iq}{2}F_{ab}\gamma^{ab}+\frac18 R_{cd;ab}\gamma^{abcd}+\frac12 \gamma^{ab}G_{ab}-\frac12\gamma^{ab}T^c_{ab}{\cal D}_c
-\frac12{R^b}_{a;db}\gamma^{ad}\label{Weitzenbock}
\eeqn
In the absence of torsion, the curvature tensors satisfy \(\lcR_{ab;cd}=\lcR_{cd;ab}\) and \({{\lcR}^b}_{\kern5pta;bd}={{\lcR} ^b}_{\kern5ptd;ba}\). Therefore the last four terms in \eqref{Weitzenbock} would vanish in the torsionless case.

\section{The Hall Viscosity} \label{section 5}

Here, we will give an extensive discussion of the Hall viscosity for the Dirac model in 2+1. We begin with 3 separate derivations (which are of course equivalent): the Berry curvature due to area-preserving diffeomorphisms, linear response via the stress-stress correlator, and the effective action in background fields. In each of these three subsections, we will obtain the same result, written in terms of an integral $I_T(m)$, which contains a power divergence at high-energies. In the follow-up section, we describe how to interpret this divergence in terms of a Pauli-Villars regularization scheme. 

\subsection{The Berry Curvature}

We begin with the direct calculation of the Berry curvature for a Dirac field on a spatial torus with complex structure $\tau$. This calculation is the analogue of that done by Avron et al. for IQHE\cite{avron1995,levay1997,read2009,tokatly2009}. 

We consider a square torus, made in $\mathbb{R}^2$ by identifications $(x,y)\sim (x+a L,y+b L)$ with $a,b\in\mathbb{Z}$. We take this to have fixed volume $L^2$, and consider area preserving diffeomorphisms, which we take to correspond to modifications of the metric of the form
\beq
g=\frac{1}{\tau_2}\begin{pmatrix}1&\tau_1\cr\tau_1 &|\tau|^2\end{pmatrix},\ \ \ \ \ 
g^{-1}=\begin{pmatrix}\frac{|\tau|^2}{\tau_2}&-\frac{\tau_1}{\tau_2}\cr-\frac{\tau_1}{\tau_2} &\frac{1}{\tau_2}\end{pmatrix}
\eeq
The basis vectors are then
\beq
\e_1=\sqrt{\tau_2}\pa_{x},\ \ \ \ \ 
\e_2=\frac{1}{\sqrt{\tau_2}}\left(-\tau_1\pa_{x}+\pa_{y}\right)
\eeq
and the veirbein is
\beq
e^1=\frac{1}{\sqrt{\tau_2}} (dx-\tau_1dy),\ \ \ \ \ 
e^2=\sqrt{\tau_2}dy
\eeq
Since the components of the metric are constant, we will take the connection to vanish. 
Take a basis for $C\ell(2,1)$, $\gamma^0=i\sigma_3$, $\gamma^1=\sigma_1$ and $\gamma^2=\sigma_2$.  The Dirac operator is
\beq
\pasl=\gamma^a \e^\mu_a\pa_\mu
=
 \begin{pmatrix} i\partial_0 &\frac{i}{\sqrt{\tau_2}}\left(\bar\tau\pa_{ x}-\pa_{ y}\right)\cr \frac{i}{\sqrt{\tau_2}}\left(-\tau\pa_{ x}+\pa_{ y}\right)& -i\partial_0\end{pmatrix}
\eeq
and the Hamiltonian is then 
\beq
H= \begin{pmatrix} m &{\bf p}\cr 
{\bf\bar p}& -m\end{pmatrix}
\eeq
where ${\bf p}=\frac{1}{\sqrt{\tau_2}}\left(\bar\tau p_x-p_y\right)$. 
The eigenstate energies thus satisfy
\beq\label{energyspectau}
E^2={\bf p\bar p}+m^2\equiv ||p||^2+m^2=\frac{|\tau|^2p_x^2-2\tau_1 p_xp_y+p_y^2}{\tau_2}+m^2.
\eeq
We introduce the phase $\eta^2=\frac{{\bf p}}{||p||}=\sqrt{\frac{p}{\bar p}}.$ 
States in the negative energy band can then be written in the form
\beq
\psi_-(p_x,p_y;\tau)=\begin{pmatrix}\eta\sqrt{\frac{|E|- m}{2|E|}}\cr -\eta^*\sqrt{\frac{|E|+m}{2|E|}}\end{pmatrix}
\eeq
The states in the upper energy band are
\beq
\psi_+(p_x,p_y;\tau)=\begin{pmatrix}\eta\sqrt{\frac{|E|+ m}{2|E|}}\cr \eta^*\sqrt{\frac{|E|-m}{2|E|}}\end{pmatrix}
\eeq
with $E$ given by (\ref{energyspectau}).
On the torus, the components of momentum are of the form
\beq
p_x=2\pi(q+h/2),\ \ \ \ p_y=2\pi(r+k/2),\ \ \ \ \ q,r\in\mathbb{Z}
\eeq
where $(h,k)$ label the spin structures.\footnote{That is we require
\[ \psi(x+a,y+b)=e^{i\pi (ha+kb)}\psi(x,y),\ \ \ \ \ a,b\in\mathbb{Z}.\]
}
We will consider the insulating ground state in which all of the negative energy states are occupied and the positive energy states are empty: 
\beq
|\Omega\rangle=\prod_{\vec k} d_{\vec k}^\dagger {\cal P}_{\vec k}(\tau)|\pm\rangle\rangle_{\vec k}
\eeq
where ${\cal P}_{\vec k}(\tau)$ is the projection operator onto the negative energy state at each momentum $\vec{k}$ for modular parameter $\tau.$ From the above considerations, we find
\beq
{\cal P}_{\vec k}(\tau)=\begin{pmatrix}\eta^*\sqrt{\frac{|E|- m}{2|E|}}&-\eta\sqrt{\frac{|E|+ m}{2|E|}}\cr
0&0\end{pmatrix}
\eeq
The Berry connection is
\beqn
A&=&-i\langle\Omega | d |\Omega\rangle\\
&=&-i\sum_{\vec k} tr {\cal P}^\dagger_{\vec k}(\tau) d{\cal P}_{\vec k}(\tau)\\
&=& i\sum_{q,r\in\mathbb{Z}}\psi^\dagger_-(q,r;\tau)d\psi_-(q,r;\tau)
\eeqn
In writing this expression, we have explicitly assumed that the oscillators have no $\tau$-dependence and are normalized via $\{d_{q,r}^\dagger,d_{q',r'}\}=\frac{i}{2}\delta_{q,-q'}\delta_{r,-r'}$. A straightforward calculation then gives
\beq
A=-i\sum_{q,r\in\mathbb{Z}} f(||p||^2) (\eta^*)^2d\left(\eta^2\right)
\eeq
where
\beq
f(||p||^2)=\frac{m}{\sqrt{m^2+||p||^2}}
\eeq
We find the Berry curvature from taking the exterior derivative
 \beqn
 F&=&i\sum_{m,n\in\mathbb{Z}}f'(||p||^2) d{\bf p}\wedge d{\bf\bar p}\\
 &=&i\frac{d\tau\wedge d\bar\tau}{2\tau_2}\sum_{q,r}p_x^2 f'(||p||^2)
 \eeqn
This discrete sum can be cast as an integral in the large volume limit 
\beq
\Sigma\equiv \sum_{q,r}p_x^2 f'(||p||^2)\to L^2\int \frac{d^2p}{(2\pi)^2} f'(||p||^2) p_x^2
\eeq
We introduce polar coordinates by transforming
\beq
\sqrt{\tau_2}p_x=||p||\cos\theta,\ \ \ \ 
\frac{1}{\sqrt{\tau_2}}(-\tau_1 p_x+p_y)=||p||\sin\theta
\eeq
so the integral becomes
\beqn
\Sigma&=&\frac{L^2}{(2\pi)^2\tau_2}\int_0^\infty dp\ p^3 f'(p^2)\int_0^{2\pi}d\theta\cos^2\theta\\
&=&
\frac{L^2}{8\pi\tau_2}\int_0^\infty dy\ y f'(y)\\
&=&\frac{L^2}{8\pi\tau_2}\int_0^\infty dy\ y \frac{d}{dy} \frac{m}{\sqrt{m^2+y}}\label{moverEintegral}\\
&=&\frac{L^2}{8\pi\tau_2}I_T(m)
\eeqn
This defines the integral $I_T(m)$ which diverges at short distances (large $||p||$). Thus, the calculation should be performed with an explicit regulator, a subject we will address carefully below. In the calculation performed by Avron \emph{et al.} for the IQHE produced from Landau levels the result is finite and proportional to $\frac{1}{\ell_{B}^2}$ where $\ell_{B}$ is the magnetic length; a short-distance cut-off generated by the uniform magnetic field. For the Dirac insulator, and topological insulators in general there is no such natural length scale which is why the calculation must be carefully regularized.

\subsection{The Stress-Stress Correlator}

Now, let us consider the 2+1 Dirac theory and look directly at $\langle T_{\mu\nu}T_{\lambda\rho}\rangle$. By general arguments, the Berry curvature computed above can be related to (the parity odd part of) this correlator, and in particular to the Hall viscosity. We note that a careful analysis of the Kubo-formula description of Hall viscosity transport, and its connection to conductivity for Galilean invariant systems, is discussed in Ref. \cite{bradlyn2012}.  In any case, we show here the explicit calculation for clarity. Recall first the Lagrangian
\beq
L=\sqrt{g}\bar\psi(t,\vec x)\left[ i\gamma^0\pa_t-m+i\sum_{i=x,y}\sum_{a=1,2}\e^i_a \pa_i\gamma^a\right]\psi(t,\vec x)
\eeq

The stress tensor is then given by
\beqn
T_{ij}(t,\vec x)&=&\ :\bar\psi(t,\vec x)\left[\eta_{ac}e^c_{(j} i\pa_{i)}\gamma^a\right]\psi(t,\vec x):+\delta_{ij}L\\
T_{0i}(t,\vec x)&=&\ \frac12 :\bar\psi(t,\vec x)\left[ i\pa_{i}\gamma^0\right]\psi(t,\vec x):
\eeqn
we compute the correlator
\beqn
\langle T_{ij}(t,\vec x)T_{k\ell}(0;\vec 0)\rangle &=& 
-4tr\left[\frac12\eta_{ac}e^c_{(j} i\pa^{x}_{i)}\gamma^a\right]iS_F(t;\vec x)
\left[\frac12\eta_{bd}e^d_{(\ell} i\pa^{x'}_{k)}\gamma^b\right]iS_F(-t;-\vec x)
\eeqn
where $iS_F=(i\pasl-m)^{-1}$. The $\delta_{ij}$ term in $T_{ij}$ will not contribute to what we are interested in (although it would contribute to other visco-elastic response coefficients). Since we are just interested in this response, we now set  $e^c_j=\delta^c_j.$ Now we can Fourier transform 
\beqn
\langle T_{ij}T_{k\ell}\rangle(p) &=& 
-4tr\int \frac{d^3q}{(2\pi)^3}\frac{\left[\frac12q_{(i}\gamma_{j)}\right](\qsl+m)
\left[\frac12r_{(k}\gamma_{\ell)}\right](\rsl+m)}{(q^2-m^2(\vec q^2))(r^2-m^2)}
\eeqn
where $\vec r\equiv\vec p+\vec q$.
The terms that have an Levi-Civita tensor are the following
\beqn
-\int \frac{d^3q}{(2\pi)^3}\frac{m}{(q^2-m^2)(r^2-m^2)}\ q_{(i}r_{(k}\ tr\ \gamma_{j)}\qsl \gamma_{\ell)}
\\
-\int \frac{d^3q}{(2\pi)^3}\frac{m}{(q^2-m^2)(r^2-m^2)}\ q_{(i}r_{(k}\ tr\ \gamma_{j)} \gamma_{\ell)}\rsl
\gamma_{\ell)}\rsl
\eeqn
This can be rewritten ($\psl=\omega\gamma^0+p_m\gamma^m$, $q^2=\omega_q^2-\vec q^2$)
\beqn
2i\epsilon_{(\ell (j}\int \frac{d^3q}{(2\pi)^3}\frac{m q_{i)}r_{k)} \omega_q}{(q^2-m^2)(r^2-m^2)}
-2i\epsilon_{(\ell (j}\int \frac{d^3q}{(2\pi)^3}\frac{m q_{i)}r_{k)}\omega_r}{(q^2-m^2)(r^2-m^2)}
\eeqn
where we used $tr\gamma^0\gamma_i\gamma_j=-2i\epsilon_{ij}$ in all terms. Now we expand this to zeroth order in the external momentum $\vec p$ since it is already first order in the external frequency, and get
\beqn
-2im \omega_p\epsilon_{(\ell (j}\int \frac{d^3q}{(2\pi)^3}\frac{q_{i)}q_{k)}}{(q^2-m^2)(r^2-m^2)}
\eeqn
Here we used $(\omega_q-\omega_r)=-\omega_p$. We can do the $\omega_q$ integral, which is trigonometric:
\beqn
&&\frac12\omega_p\int \frac{d^2q}{(2\pi)^2}\frac{m\epsilon_{(\ell (j}q_{i)}q_{k)}
}{(\vec q^2+m^2)^{3/2}}\\
&=&-\frac{1}{4}\omega_p\epsilon_{(\ell (j}\delta_{i)k)}\int \frac{d^2q}{(2\pi)^2}\vec q^2\frac{\pa}{\pa\vec q^2}\frac{m
}{(\vec q^2+m^2)^{1/2}}\nonumber\\
&=&-\frac{1}{16\pi}\omega_p\epsilon_{(\ell (j}\delta_{i)k)}I_T(m)
\eeqn
This is the same integral that appears in the Berry phase calculation.
One can also similarly establish
\beqn
\langle T_{0i}T_{jk}\rangle(p) &=& 
\frac{1}{32\pi}p_n\delta_{i(j}\epsilon_{k)n}\int_0^\infty dy\ y\frac{\pa}{\pa y}\frac{m(y)
}{(y+m^2(y))^{1/2}}
\eeqn
and thus for small $p^\mu$, we have
\beqn
\langle T_{\mu\nu}T_{\lambda\rho}\rangle(p) &=& 
\frac{1}{16\pi}\delta_{(\nu(\rho}\epsilon_{\mu)\lambda)\sigma}p^\sigma I_T(m)\label{eq:TTviscosity}
\eeqn

\subsection{The Background Field Formalism} \label{subsec: hall viscosity}

This result in Eq. \ref{eq:TTviscosity} can also be rewritten in terms of the stress current (as above, a vector-valued 1-form current associated with diffeomorphisms)
\beqn
\langle J^a_\mu J^b_\lambda\rangle(p) &=& 
\frac{1}{16\pi}\eta^{ab}\epsilon_{\mu\lambda\sigma}p^\sigma I_T(m)+\ldots
\eeqn
At leading order, this may be interpreted as coming from a quadratic parity-violating term in the effective background-field action of the form 
\beq
S_{odd}[e,\omega,A]=\left(...+\frac{1}{32\pi}I_T(m,\Lambda)\int e^a\wedge de^b\eta_{ab}+...
\right)
\eeq
In addition to this term, there will be a higher order term that renders\footnote{In a generic state of a condensed matter system, we of course have no reason to suppose that Lorentz invariance would be obtained. However, the specific model under discussion does have local Lorentz covariance, and thus the effective action will share that feature.} it locally Lorentz invariant
\beq\label{effacttorsion}
S_{odd}[e,\omega,A]=\left(...+\frac{1}{32\pi}I_T(m)\int e^a\wedge T^b\eta_{ab}+...
\right)
\eeq
This result means that the Dirac fermion contributes $\frac{1}{16\pi}I_T(m)$ to the Hall viscosity $\zeta_H$. Again, we stress that the result is divergent, but we will consider this divergence carefully later in the context of a Pauli-Villars regulator.

It is instructive to produce (\ref{effacttorsion}) directly via a background field expansion. Here what we will do is consider the Dirac fermion in a general background co-frame $e^a$ and connection ${\omega^a}_b$ and integrate out the fermion. This is in principle straightforward, since the fermion integral is Gaussian and the fermion is gapped, although the result is non-standard, as we will consider a torsional connection. Later we will extend the calculation to include the effects of curvature, as well as a calculation of the \emph{parity-even} terms.

We have written the general Dirac action in a previous section. Here we are interested in computing the determinant of the Dirac operator (now for a general value of $\torcplg$)
\beqn
\cDsl&=&\gamma^ae_a^\mu\left(\pa_\mu+A_\mu^At_A+\frac14 \omega_{\mu;bc}\gamma^{bc}\right)+B_b\gamma^b+\frac18\alpha T_{a;bc}\gamma^{abc}\\
&=&\cDslLC+\frac18\torcplg T_{a;bc}\gamma^{abc}
\eeqn
where $B_a\equiv \frac{1}{2}T^b_{ab}$. In particular, the contribution to the effective action is $-\ln\det (\cDsl-m)$. Formally, we write this as
\beqn
-\ln\det (\cDsl-m)&=&-\ln\det({\cDslLC-m})-\ln\det\frac{\cDsl-m}{\cDslLC-m}\\
&=&-\ln\det({\cDslLC-m})-\ln\det\frac{\cDsl-m}{\cDslLC-m}\frac{\cDslLC+m}{\cDslLC+m}\\
&=&-\ln\det({\cDslLC-m})-\ln\det\frac{\cDslLCsq-m^2+(\cDsl-\cDslLC)(\cDslLC+m)}{\cDslLCsq-m^2}
\eeqn
where by $\cDslLC$, we mean the Dirac operator with (torsionless) Levi-Civita connection $\stackrel{\circ}{\omega}$ , while $\cDsl$ is a general torsional connection. The second expression vanishes in the absence of torsion while $\ln\det (\cDslLC-m)$ is the usual gravitational effective action (which will contain a volume divergence, the gravitational and gauge Chern Simons terms, as well as higher order terms). See Ref. \cite{Ojima} for related calculations.

Let us then focus on the second expression. 
One advantage to this quantity is that the volume divergence (being independent of the connection) has to cancel exactly, because the numerator and denominator differ only in the connection (and not the frame). The expression is well-defined mathematically, as it is comparing the contributions of two different connections (analogue of  relative Chern-Simons). 

We have (as a special case of \eqref{Weitzenbock})
\beqn
\cDslLCsq&=&
\eta^{ab}\cDLC_a\cDLC_b-\frac14 \RLC+\frac{iq}{2}F_{ab}\gamma^{ab}
\eeqn
and one can establish \footnote{Here we have a choice for the representation of the Clifford algebra, namely $\gamma^{abc} = \pm\epsilon^{abc}$, which essentially amounts to a choice of orientation. We will choose the positive sign; this will be reflected in the sign of $\zeta_H$, and later on other parity odd coefficients.}

\beqn
\cDsl-\cDslLC&=&\frac14C_{bc}(e_a)\gamma^a\gamma^{bc}+B_b\gamma^b+\frac18\alpha T_{a;bc}\gamma^{abc}\\
&=&\frac{\torcplg }{4}C_{bc}(e_a)\gamma^{abc}=\frac{\torcplg }{4} \epsilon^{abc}C_{a;bc}  \equiv \frac14\torcplg  c
\eeqn

Formally, we may write
\beqn
-\ln\det\frac{\cDsl-m}{\cDslLC-m}&=&-\ln\det \left( 1 + \frac{\torcplg }{4} c (\cDslLC-m)^{-1}\right)\\
&\simeq&-\frac{\torcplg }{4} tr\ c(\cDslLC-m)^{-1}+...
\eeqn
So the entire effect of torsion can be thought of in terms of a $c\bar\psi\psi$ vertex, and the above expression just corresponds to the fermion loop diagram with a single insertion of $c$ (the singlet part of the contorsion).
It will be sufficient to work to first order in $c$ to see the torsion Chern-Simons term, as
\beq
e^a\wedge C_{ab}\wedge e^b=e^a\wedge T_a=d^3x\ \varepsilon^{\mu\nu\lambda}e^a_\mu C_{\nu;ab} e^b_\lambda =- c d^3x
\eeq
where in the last equality we expanded around flat space-time. 
Indeed, to extract just the torsional CS term, we can expand around flat space-time to obtain the leading result
\beqn
-\frac12m\;\mathrm{Tr}\ c(\Box-m^2)^{-1}&=&\frac12m\int d^3x\ c(x)\int\frac{d^3k}{(2\pi)^3}\frac{1}{k^2+m^2} \label{flatexp}\\
&=&
\frac12im\int d^3x\ c(x)\int_\epsilon^\infty ds  \int\frac{d^3k}{(2\pi)^3}e^{-i(k^2+m^2)s}\\
&=&
\frac12imI(0,0;|m|)\int d^3x c(x)
\eeqn

Here we are using the notation
\beq
I(p,q;|M|)=\int \frac{ds}{s^p}\int \frac{d^3k}{(2\pi)^3}e^{-is(k^2+M^2)}|k|^q \label{I(p,q)}
\eeq
Now we can do a Wick rotation in the complex $k_0$ plane ($k_0=-ik_4$) and then rotate the contour in the complex $s$-plane ($s=-it$) to get
\beq
I(p,q;|M|)=-i^{p}\int \frac{dt}{t^p}\int \frac{dk}{2\pi^2}e^{-t(k^2+M^2)}k^{q+2}
\eeq
We note that the $k$-integral converges in the UV unless $t=0$. So we cut off the $t$-integration at $t=\epsilon\sim 1/\Lambda^2$.
\beqn
I(p,q;M)
&=&-\frac{i^{p}}{4\pi^2}\Gamma\left(\frac{q+3}{2}\right)|M|^{2p+q+1}\Gamma\left(-p-\frac{q}{2}-\frac{1}{2},\epsilon |M|^2\right)
\eeqn
In terms of the previous notation, one finds $I_T(m)=8\pi m I(0,0;|m|)$. For later use, we note that for small $\epsilon$, we have
\beqn
iI(-1,0;m)&=&-\frac{1}{8\pi |m|}+...\label{expand1}\\
I(0,0;m)&=&-\frac{1}{2\pi}\frac{1}{\sqrt{4\pi\epsilon}}+\frac{|m|}{4\pi}+...\label{expand2}\\
iI(1,0;m)&=&\frac{2}{3}\frac{1}{(4\pi\epsilon)^{3/2}}-\frac{|m|^2}{2\pi}\frac{1}{\sqrt{4\pi\epsilon}}+\frac{|m|^3}{6\pi}+...\label{expand3}
\eeqn

There is also a Chern-Simons term for the \(U(1)\) gauge field, including which we get
\beq
S_{odd}[e,\omega,A]=-\frac{i}{2} \torcplg mI(0,0;|m|)\int e^a\wedge T_a +q^2\; mI(-1,0;|m|) \int A\wedge dA 
\eeq
Comparing these results with \eqref{CStorsionfree} and \eqref{CStorsional}, we get the unregulated coefficients for a single Dirac fermion of mass $m$ and charge $q$
\beqn
\zeta_H(m)&=&-\torcplg m I(0,0;|m|)\\
\sigma_H(m)&=& - 2i q^2\;m I(-1,0;|m|)
\eeqn



\subsection{Pauli-Villars Regulator}

Now, we will finally consider a proper regularization for the Hall conductivity and Hall viscosity. To do this we introduce a set of $N$ regulator fields with masses $M_i$ and normalizations $C_i$. The regulated coefficients are then
\beqn
\zeta_H(\{M_i\})&=&-\torcplg \sum_{i=0}^N C_i  M_i I(0,0;|M_i|)\\
\sigma_H(\{M_i\})&=&-2iq^2 \sum_{i=0}^N C_i  M_i I(-1,0;|M_i|)
\eeqn
where the original fermion is labeled with $M_0=m$ and $C_0=1$. 

What we should do here is choose the remaining $C_i$ and $M_i$ such that the resulting expressions are finite as $\epsilon\to 0$. Making use of the expansions (\ref{expand1}--\ref{expand2}), we find as $\epsilon\to 0$
\beqn
\zeta_H(\{M_i\})&=&-\frac{\torcplg }{4\pi}\sum_{i=0}^N C_i  \left[-\frac{M_i}{\sqrt{\pi\epsilon}}+\sigma_i|M_i|^2+...\right]\\\label{regcoeff1}
\sigma_H(\{M_i\})&=&\frac{q^2}{4\pi} \sum_{i=0}^N C_i  \sigma_i+... \label{regcoeff2}
\eeqn
where $\sigma_i\equiv sign(M_i)$, and in particular $\sigma_0=sign(m)$.
To render these expressions finite, we must require
\beqn\label{finiteness}
\sum_{i=0}^N C_i M_i=0.
\eeqn
We note that in fact this condition must \emph{always} be satisfied regardless of whether or not the system is a trivial or a topologic insulator, that is, it is satisfied independent of the sign of $m.$ To see that this can be achieved, we take as a guide the (2+1)-d lattice Dirac model\cite{qi2008b} with Hamiltonian
\beq
H=\sum_{\vec k;a,b}= c_{\vec k;a}^\dagger \left[ (m+\mu_{bw}(\cos k_x+\cos k_y-2))\sigma_3 +v_F\sin k_x \sigma_2+v_F\sin k_y\sigma_1
\right]_{ab}c_{\vec k;b}
\eeq  We compare these continuum calculations to the lattice Dirac model which can be interpreted as having one low-energy fermion and three regulator fermions.
On the lattice, momenta are measured in units of the lattice spacing $a$, and the Brillouin zone is a torus given by the square $\left[-\frac{\pi}{a},\frac{\pi}{a}\right]\times\left[-\frac{\pi}{a},\frac{\pi}{a}\right]$ with the boundaries identified. The spectrum 
can be interpreted as four independent copies of the (2+1)-d continuum Dirac Hamiltonian located at four separate lattice momenta $\vec k=(0,0), (0,\frac{\pi}{a}), (\frac{\pi}{a},0), (\frac{\pi}{a},\frac{\pi}{a})$. The Dirac fermions away from the origin can be thought of as analogous to the Pauli-Villars regulator fields, and the regulated continuum model is then an approximation to the lattice model where only the effects of these Dirac points are taken into account.
 Examination of the Hamiltonian near these points gives

\begin{center}\begin{tabular}{|c|c|}
\hline
 $M_i$ & $C_i$ \\
 \hline
$m$   & +  \\
$m-2\mu_{bw} $  & -   \\
$m-2\mu_{bw} $  & -   \\
$m-4\mu_{bw}$   & +  \\
\hline
\end{tabular}\end{center}

and we note that indeed $\sum_{i=0}^N C_i M_i=0$, independent of the value of $m$.

Returning to the general Pauli-Villars problem then, we will take $N=3$ with\footnote{Since each of the Dirac points has a Hamiltonian linear in momentum, $C_i$ refers to the ``parity" of the Dirac copy; thus the modes at $(0,\pi), (\pi,0)$ have opposite parity to the other two. Parity can be understood as the relative sign between the two terms in the Dirac Hamiltonian which are linear in momentum. In the following section, we will see that the condition $\sum_{i=0}^N C_i=0$ will follow from finiteness in the parity even sector. This is the usual ``fermion doubling" phenomenon.} $C_i=\{+,-,-,+\}$ and then take masses such that $M_1+M_2-M_3=m$, say by taking $M_1=m+\Lambda_1$, $M_2=m+\Lambda_2$, $M_3=m+\Lambda_1+\Lambda_2$, where we are considering $m<<\Lambda_{1,2}$. Now, if we use $\sigma_{H}$ as a guide, we believe that the sign of $m$ determines two phases, one of which is a trivial insulator and one of which is a non-trivial topological insulator.
A choice of the signs of $\Lambda_{1,2}$ will correspond to choosing which value of $\sigma_0=sign(m)$ gives the trivial phase. For definiteness then, let us choose $\Lambda_{1,2}>0$. We thus obtain
\beqn
\zeta_H&=&-\frac{\torcplg}{4\pi}\left[\sigma_0 |m^2| -(m+\Lambda_1)^2-(m+\Lambda_2)^2+(m+\Lambda_1+\Lambda_2)^2\right]\nonumber\\
&=&- \frac{\torcplg }{4\pi}\left[\sigma_0 |m|^2 -|m|^2+2\Lambda_1\Lambda_2\right]\\
\sigma_H&=&\frac{q^2}{4\pi}\left[ \sigma_0-1 -1+1\right]
\eeqn
Thus we see that
\beqn
\left\{ \begin{tabular}{ll}
$m>0$:& \begin{tabular}{l} $\zeta_H = \frac{\torcplg }{2\pi}\Lambda_1\Lambda_2$ \\ $\sigma_H=0$\end{tabular} 
\\ \\
$m<0$:& \begin{tabular}{l} $\zeta_H = \frac{\torcplg }{2\pi}\Lambda_1\Lambda_2+\frac{\torcplg }{2\pi}m^2$ \\ $\sigma_H=-\frac{q^2}{2\pi}$\end{tabular}
\end{tabular}
\right.
\eeqn
The value of $\zeta_H$ diverges as we remove the regulator masses to infinity, but in a way that is independent of the phase. Thus, including a counterterm to subtract this divergence (or by including additional regulator fields appropriately), we see that (the renormalized) coefficients $\zeta_H$ and $\sigma_H$  vanish in the trivial phase, while in the non-trivial phase, they are
\beq
\sigma_H= -\frac{q^2}{2\pi},\;\;\zeta_H=\torcplg \frac{m^2}{2\pi}\label{coeffs}
\eeq
Said another way, it is the {\it difference} of $\zeta_H$ (as well as $\sigma_H$) in the two phases that is universal. In fact, this difference can be detected by studying anomalous conservation laws for 1+1 dimensional chiral fermions which live on the edge separating the trivial and non-trivial phases, as we will see shortly. We give a more physical discussion of the Hall viscosity in terms of the bulk-boundary correspondence in Section \ref{sec:spectralflow}.

\section{Effective action as Chiral gravity} \label{sec: chiral gravity}

In the previous section, we have shown how to extract and regulate the Hall viscosity term, working in a flat background.  In this section, we compute the effective action for massive Dirac fermions coupled to a generic background frame and connection. 
We will work in the limit of large radii of curvature and torsion in comparison with the fermion correlation length \(\xi \sim \frac{1}{m}\).  We will utilize the asymptotic expansion for the trace of the heat kernel of \(\slashed{\msD}^2\) (see Appendix \ref{sec: AE}) to extract terms which survive as \((ma)\rightarrow \infty\), \(a\) being the radius of curvature. The resulting renormalized action in the non-trivial phase differs from the trivial phase by an \(SL(2,\mathbb{R})\) Chern-Simons term.

We will work on a manifold $M$ with Euclidean signature. At this point, we might expect that the action can be written in the form
\beqn
\frac12\int\Big(i\sigma_H\;A\wedge dA +i \kappa_HCS[\omega]+i\zeta_H\; e^a\wedge T_a  +\frac{1}{\kappa_N}\epsilon_{abc}e^a\wedge R^{bc}-\frac{2\Lambda}{\kappa_N}vol_M +...
\Big)
\label{EinsteinHilbert}
\eeqn
where $CS[\omega]= \omega_{ab}\wedge d\omega^{ba}+\frac23{\omega^a}_b\wedge {\omega^b}_c\wedge {\omega^c}_a$ and $vol_M=\frac{1}{3!}\epsilon_{abc}e^a\wedge e^b\wedge e^c$.  More precisely, we will eventually find that the leading terms in the (renormalized)  action will organize into a specific form involving $e^a$, $\lcw^{ab}$ and $H$.

\subsection{Parity odd terms}
In the absence of background torsion, the parity odd effective action for the Dirac theory is given by \cite{Ojima, AGPM}\footnote{We will assume for convenience that \(\cDslLC\) and \(\slashed{\msD}\) have no zero modes.}
\beq
S_{odd}[e,\lcw,A]=mI(-1,0;|m|) q^2\int A\wedge dA + \frac{1}{24}mI(-1,0;|m|)\int \mathrm{tr}(\LCconn\wedge d\LCconn+\frac{2}{3}\LCconn\wedge \LCconn\wedge \LCconn)+...
\eeq
where the integral \(I(p,q; |m|)\) has been defined in \eqref{I(p,q)}. In order to extract terms to leading order in torsion, the simplest thing we can do is repeat the relative Chern-Simons calculation of the previous section in the presence of background curvature. We thus replace \eqref{flatexp} with 
\beqn
S_{odd}[e,\omega,A]&=&S_{odd}[e,\lcw,A]+\frac{i}{4}\torcplg m\;\mathrm{Tr}\;c(-\cDslLC^2 + m^2)^{-1} +...\nonumber\\
&=& S_{odd}[e,\lcw,A]+\frac{i}{4}\torcplg m\int_0^{\infty} dt\;\mathrm{Tr}\;c e^{-t(-\cDslLC^2 + m^2)} +...\label{relativeCScurved}
\eeqn
where as before, \(c=\epsilon^{abc}C_{a;bc}\). In the limit \(t\rightarrow 0\), there exists an \textsl{asymptotic expansion} for \(\mathrm{Tr}\; e^{t\slashed{\msD}^2}\) (in arbitrary dimension \(d\))
\beq
\mathrm{Tr}\; e^{t\slashed{\msD}^2} \simeq \frac{1}{(4\pi t)^{d/2}} \sum_{k=0}^{\infty} a_k t^{k}
\eeq
where \(a_k\) are integrals over \(M\) of polynomials in curvature, torsion and their covariant derivatives. The important point is that  it suffices to use this asymptotic expansion in order to extract terms which survive as \((ma)\rightarrow \infty\). For the case at hand, we have (see Appendix \ref{sec: AE} for details)
\beq
\mathrm{Tr}\;e^{t\cDslLC^2} \simeq \int\frac{2}{(4\pi t)^{3/2}}\;\left(1-\frac{t}{12}\lcR + O(t^2) \right)vol_M \label{LCheatkernel3}
\eeq
where \(vol_M\) is the volume form on \(M\). The \(O(t^2)\) terms are unimportant for the present discussion, as they are higher order in powers of curvature and lead to negative powers of \((ma)\). From \eqref{relativeCScurved} and \eqref{LCheatkernel3}, we get 
\beqn
S_{odd}[e,\omega,A] &=& -\frac{i}{2}\torcplg mI(0,0;|m|) \int e^a\wedge T_a+mI(-1,0;|m|) q^2\int A\wedge dA \ \label{parityodd1}\\
&+& \frac{1}{24}mI(-1,0;|m|)\int \left(\mathrm{tr}\;(\LCconn\wedge d\LCconn+\frac{2}{3}\LCconn\wedge \LCconn\wedge \LCconn)-\torcplg \lcR\;e^a\wedge T_a+\cdots\right)\nonumber
\eeqn
where the ellipsis indicates terms higher order in torsion. 

It will also be convenient to introduce a new \(SO(3)\) connection \(\omega^{(\beta)}_{\mu;ab}\equiv\lcw_{\mu;ab}-\frac{\beta}{2}H_{\mu;ab}\). This is natural because, as we have seen, the Dirac operator involves only $\lcw$ and $H$.  If we choose the specific value $\beta=-\torcplg$, one finds that the second line of \eqref{parityodd1} combines into a single Chern-Simons term for $\omega^{(-\torcplg)}_{\mu;ab}$.
 In fact, this is indicated by the structure of the chiral anomaly in \(d=4\) (see Eq \eqref{4dchiralanomaly} in Appendix \ref{sec: AE}).\footnote{The connection  between the \(d=2n+1\) parity odd effective action and \(d=2n\) chiral anomaly is provided by the Atiyah-Patodi-Singer index theorem; see \cite{AGPM} for details.} Given the level of the calculation that we have presented, it is possible to verify this to linear order in torsion. Equation \eqref{parityodd1} may be rewritten in these terms (to linear order in torsion) as
\beqn
S_{odd}[e,\omega,A] &=& -\frac{i}{2}\torcplg mI(0,0;|m|) \int e^a\wedge T_a+q^2mI(-1,0;|m|) \int A\wedge dA \label{parityodd3}\\
&+& \frac{1}{24}mI(-1,0;|m|)\int \mathrm{tr}\;(\omega^{(-\torcplg)}\wedge d\omega^{(-\torcplg)}+\frac{2}{3}\omega^{(-\torcplg)}\wedge \omega^{(-\torcplg)}\wedge \omega^{(-\torcplg)}) \nonumber
\eeqn

\subsection{Parity even terms}

The parity even terms in the effective action are given by
\beq
S_{even}[e,\omega,A]= \lim_{\epsilon \to 0^+}\int_{\epsilon}^{\infty}\frac{dt}{2t}\mathrm{Tr}\;e^{-tm^2+t\slashed{\msD}^2}\label{evendefn}
\eeq
Once again, it suffices to use the asymptotic expansion for \(\mathrm{Tr}\;e^{t\slashed{\msD}^2}\) in order to compute terms which survive in the large \((ma)\) limit. The asymptotic expansion in this case is given by (see Appendix \ref{sec: AE})
\beq
\mathrm{Tr}\;e^{t\slashed{\msD}^2} \simeq \int\frac{2}{(4\pi t)^{3/2}}\;\left(1-\frac{t}{12}R^{(-\torcplg)} + O(t^2) \right)vol_M \label{heatkernel4}
\eeq
where \(R^{(-\torcplg)}=\lcR-\frac{\torcplg^2}{4}H_{abc}H^{abc}\) is the scalar curvature constructed out of \(\omega^{(-\torcplg)}_{ab}\). Using \eqref{evendefn} and \eqref{heatkernel4}, we get
\beqn
S_{even} &=& \lim_{\epsilon \to 0^+}\int  \left(iI_{\epsilon}(1,0;|m|)+I_{\epsilon}(0,0;|m|)\frac{R^{(-\torcplg)}}{12}+\cdots\right)vol_M\\
&=& \lim_{\epsilon \to 0^+}\int  \left(iI_{\epsilon}(1,0;|m|)vol_M-\frac{1}{12}I_{\epsilon}(0,0;|m|)\epsilon_{abc}e^a\wedge R^{(-\torcplg),bc}+\cdots\right)
\eeqn
where the ellipsis indicates terms of order \((ma)^{-1}\). 
\newcommand{\sn}{sign}


In order to regulate the divergences in the effective action, we  introduce \(N\) Pauli-Villars regulator fermions as we have done previously. Using the expansions (\ref{expand1}--\ref{expand3}), we find
\begin{equation}
\zeta_H (\{M_i\}) = -\frac{\torcplg }{4\pi}\sum_{i=0}^NC_i\left[-\frac{M_i}{\sqrt{\pi\epsilon}}+\sigma_i|M_i|^2+\cdots\right]\label{regcoefffull1}
\end{equation}
\beq
\sigma_H(\{M_i\}) = \frac{q^2}{4\pi}\sum_{i=0}^NC_i\sigma_i+\cdots\label{regcoefffull2}
\eeq
\beq
i\kappa_H (\{M_i\})= \frac{1}{96\pi}\sum_{i=0}^NC_i\sigma_i+\cdots\label{regcoefffull3}
\eeq
\beq
\frac{1}{2\kappa_N}(\{M_i\})=\frac{1}{48\pi}\sum_{i=0}^NC_i\left[-\frac{1}{\sqrt{\pi\epsilon}}+|M_i|+\cdots\right]\label{regcoefffull4}
\eeq
\beq
-\frac{\Lambda}{\kappa_N}(\{M_i\})= \sum_{i=0}^N C_i\left[\frac{2}{3(4\pi\epsilon)^{3/2}}-\frac{M_i^2}{4\pi\sqrt{\pi\epsilon}}+\frac{|M_i|^3}{6\pi}+\cdots\right]\label{regcoefffull5}
\eeq
We require that the terms that diverge as $\epsilon\to 0$ have zero coefficients. This implies
\beq\label{renormfull}
\sum_{i=0}^NC_i=0,\ \ \ \ \sum_{i=0}^NC_iM_i=0,\ \ \ \ \ \sum_{i=0}^NC_i |M_i|^2=0
\eeq
Thus, we have one new condition from the parity even sector and we now see that the first condition (we used this above) is also required by the parity even sector. If we assume for simplicity that all of the regulator masses are positive,\footnote{This assumption leads to the $m>0$ phase being trivial. Another choice would make the $m<0$ phase trivial.} then we arrive at
\beqn
\zeta_H &=&\torcplg \frac{m^2}{2\pi} \frac{1-\sigma_0}{2}\\\
\sigma_H &=& -\frac{q^2}{2\pi}\frac{1-\sigma_0}{2}\\
\kappa_H&=&- \frac{1}{48\pi}\frac{1-\sigma_0}{2}\\
\frac{1}{\kappa_N}&=&\frac{|m|}{12\pi}\frac{1-\sigma_0}{2}\\
\frac{\Lambda}{\kappa_N} &=&-\frac{1}{6\pi}\sum_{i=0}^N C_i|M_i|^3\label{chiralgravitycoeffs}
\eeqn
where again, $\sigma_0\equiv sign(m)$.
If we examine the conditions (\ref{renormfull}), we can furthermore establish that 
\beqn
\frac{\Lambda}{\kappa_N} &=&\Lambda_0^3-\frac{1}{3\pi}|m|^3\frac{1-\sigma_0}{2}
\eeqn
for a quantity $\Lambda_0$ that generally scales with the regulator masses, but is independent of $\sigma_0$.
We thus see that apart from the $\Lambda_0^3$ term, all of these coefficients vanish in the trivial phase ($\sigma_0=1$) and the effective action there is just $S_{eff}^{+}=\Lambda_0^3\int vol_M$,  a pure cosmological term (of course, there are also higher order terms in curvature and torsion, which decay exponentially or as negative powers of \((ma)\), that we have not included here; those terms then determine the transport properties of the trivial phase). The non-trivial phase has an action consisting of the same $\Lambda_0^3$ volume term, plus an action that is known as {\it chiral gravity} (as well as the usual $U(1)$ gauge Chern-Simons term). In other words, the difference of the gravitational actions between the two phases can be written in terms of the Chern-Simons form of a single $SL(2,\mathbb{R})$ connection.\footnote{Here we are using the language of real time. The group theory involved here is that the isometry group of $AdS_3$ is $\sim SO(2,2)\sim SO(2,1)\times SO(2,1)\sim SL(2,\mathbb{R})\times SL(2,\mathbb{R})$. We warn the reader that the detailed calculations are presented here in Euclidean signature. See Ref. \cite{Witten} for details.} Let us now see how this works.

In 3 dimensions, a connection one-form \(\omega_{ab}\) may be re-written as $\omega_a=\frac{1}{2}{\epsilon}_{abc}\omega^{bc}$. Constructing \(\omega_a^{(\beta)}\) this way from \(\omega_{bc}^{(\beta)}\), we may define the two \emph{chiral} connections ${\cal A_{\pm}}^a=\omega^{(\beta),a}\pm i\frac{1}{\ell} e^a$. One finds 
\beqn
iCS[{\cal A}_{\pm}^a]&=& 2i\;\left({\cal A}^a_{\pm}\wedge d{\cal A}_{\pm,a} -\frac{1}{3}\epsilon_{abc}{\cal A}_{\pm}^a\wedge {\cal A}_{\pm}^b\wedge {\cal A}_{\pm}^c\right)\\
&=&-iCS[\omega^{(\beta)}_{ab}]\mp\frac{4}{\ell^3} vol_M\mp\frac{2}{\ell}\epsilon_{abc}e^a\wedge R^{(\beta),bc}-i\frac{2}{\ell^2}e^a\wedge T^{(\beta)}_a\\
&=&-iCS[\omega^{(\beta)}_{ab}]\mp \frac{4}{\ell^3} vol_M
\mp\frac{2}{\ell}\epsilon_{abc}e^a\wedge R^{(\beta),bc}-i\frac{6\beta}{\ell^2}e^a\wedge T_a \label{SL(2,R)CS}
\eeqn
The leading terms of a generic gravitational action in 3 dimensions may be written as \cite{Witten}
\beq
S_{grav} = \frac{ik_{+}}{4\pi}CS[\mathcal{A}_{+}]-\frac{ik_{-}}{4\pi} CS[\mathcal{A}_{-}]. \label{gravaction}
\eeq
Comparing our fermion effective action with equations \eqref{SL(2,R)CS} and \eqref{gravaction}, we see that if we identify $\beta=-\torcplg$ as above, $\ell=(2|m|)^{-1}$ and $\Lambda=-1/\ell^2$, the action in the non-trivial topological phase is
\beq\label{actionjump}
S_{eff}^{-}=S_{eff}^{+}-\frac{ik_{-}}{4\pi}CS[{\cal A}_{-}^a]-\frac{iq^2}{4\pi}CS[A].
\eeq
The Chern-Simons levels evaluate to $(k_{+},k_{-})=(0,-1/24)$ - hence the term \emph{chiral gravity}.\footnote{This result satisfies the quantization condition \(k_{\pm} \in \frac{1}{48}\mathbb{Z}\) given in \cite{Witten} for manifolds which admit a spin-structure. Here we get twice that result, because we have a full Dirac fermion in 3d.} Incidentally, chiral gravity has been studied in the context of holography\cite{Strominger}, in which the gravitational fields are dynamical. Indeed if we introduce the notation \(\mu=\frac{1}{2\kappa_N\kappa_H}\) (here $\mu\ell=-1$), the Brown-Henneaux formula in asymptotically-$AdS_3$ geometries gives the central charges of the dual $1+1$-dimensional theory as
\beq
c_L = \frac{12\pi\ell}{\kappa_N}\left(1-\frac{1}{\mu\ell}\right), \;c_R=\frac{12\pi\ell}{\kappa_N}\left(1+\frac{1}{\mu\ell}\right).
\eeq
Thus  in the holographic case, we have $c_L=1, c_R=0$. In the present case, this $1+1$-dimensional matter is supported on the interface between the topological insulator phase and the trivial phase.
In the following sections, we will investigate this further and show that the parity odd transport coefficients of (\ref{actionjump}) are encoded in anomalies of the $1+1$-dimensional theory on the interface.  This will include not only the chiral anomaly, relevant to the charge sector, but also diffeomorphism and Lorentz anomalies for the gravitational sector.
We will see that indeed (\ref{actionjump}) implies that $c_L-c_R=1$.

\section{The Callan-Harvey Anomaly inflow} \label{section 7}
Consider the non-trivial phase labelled by non-vanishing parity odd coefficients \((\sigma_H, \zeta_H, \kappa_H)\) on a 2+1 dimensional manifold \(M\), separated from the trivial phase by the 1+1 dimensional boundary  \(\Sigma=\partial M\). This can be thought of in terms of a 2+1 dimensional Dirac fermion with mass \(m<0\) on \(M\), and \(m>0\) outside, with some interpolation region,  the interface \(\Sigma\),  which we refer to as the domain wall. In general, there could be multiple fermions with mass domain walls along \(\Sigma\), and their number decides \((\sigma_H, \zeta_H, \kappa_H)\). The domain wall hosts 1+1 dimensional chiral fermions, whose anomalies will encode the shifts in \((\sigma_H, \zeta_H,\kappa_H)\) between opposite sides of the domain wall\cite{callanharvey}. In the absence of curvature (we will return to the general case later), the parity odd effective action can be taken to be \footnote{The gravitational Chern Simons terms (proportional to \(\kappa_H\)) lead to currents which are proportional to the Levi-Civita scalar curvature. Hence we ignore these terms temporarily.}$^{,}$\footnote{In this section, we will use upper case letters for Lorentz indices in the bulk and lower case letters for Lorentz indices on the boundary/domain-wall \(\Sigma\).}
\begin{equation}\label{noReffact3d}
S_{odd,bulk} [e,\omega, A] = \frac{\zeta_H}{2}\int_M e^A\wedge T_A + \frac{\sigma_H}{2} \int_M A\wedge dA
\end{equation}
Let us first focus on the gauge Chern-Simons term and review its relationship with anomalies in the boundary.
In the presence of a boundary, the \(U(1)\) Chern-Simons term is diffeomorphism and Lorentz invariant, but not gauge invariant. Under a gauge transformation we have
\begin{equation}
\delta_{\alpha} S_{odd,bulk} = \frac{\sigma_H}{2}\int_M d\alpha \wedge F = \frac{\sigma_H}{2}\int_{\Sigma}  \alpha F
\end{equation}
Gauge invariance implies that this should be accounted for by the \(U(1)\) anomaly of chiral fermions localized on \(\Sigma\). For \(n_L\) left-handed and \(n_R\) right-handed chiral fermions of charge $q$ on the edge, the anomaly is given by
\begin{equation}
\delta_{\alpha}S_{\Sigma} = \frac{n_L-n_R}{4\pi}q^2\int_{\Sigma} \alpha F \label{consanomaly}
\end{equation}
This cancels the variation of the bulk action provided \(q^2(n_L-n_R) = -2\pi\sigma_H\), which is indeed the case as can be checked by constructing the localized zero modes of the bulk Dirac operator (see \cite{callanharvey} for details). The anomaly in \eqref{consanomaly} is called a \textsl{consistent} anomaly, because it is obtained by the variation of the chiral effective action in 1+1 dimensions. We refer to the corresponding non-conserved boundary charge current as \(J_{cons}\), with the anomalous Ward identity
\begin{equation}
d*J_{cons} = \frac{n_R-n_L}{4\pi}q^2\;F = \frac{\sigma_H}{2}\;F
\end{equation}
Returning to the bulk, the variation of the effective action with respect to the gauge field determines the bulk charge current
\begin{equation}
\delta S_{odd,bulk} = \sigma_H\int_M \delta A\wedge F +\frac{\sigma_H}{2} \int_{\Sigma} \delta A\wedge A\label{CSvar}
\end{equation}
We can read off the bulk charge current from here
\begin{equation}
*J_{bulk} = \sigma_H\; F
\end{equation}
which is conserved by virtue of the Bianchi identity, i.e. \(d*J_{bulk} = 0\). However, the flux of the bulk current into $\Sigma$ is non-trivial and is given by
\begin{equation}
\Delta Q_{\Sigma} = \int_{\Sigma}*J_{bulk}=\sigma_H \int_{\Sigma} F=\frac{n_R-n_L}{2\pi}q^2 \int_{\Sigma} F\label{inflow1}
\end{equation}
We can interpret this as the charge injected into the edge from the bulk, but notice that it is twice as much as the consistent anomaly in \eqref{consanomaly}. To explain this apparent discrepancy, notice from \eqref{CSvar} that there is an additional boundary current induced from the bulk
\begin{equation}
*j = \frac{\sigma_H}{2} A
\end{equation}
This prompts us to define the net boundary current \(J_{cov} = J_{cons} + j\), which we will call the \textsl{covariant current}. The conservation equation for \(J_{cov}\) is now
\begin{equation}
d*J_{cov} = \sigma_H\;F= \frac{n_R-n_L}{2\pi} F \label{covanomaly}
\end{equation}
which agrees with \eqref{inflow1}. The anomaly in the form \eqref{covanomaly} is called the \textsl{covariant anomaly}.\footnote{The reason for the terminology \textsl{consistent} and \textsl{covariant} comes from the more general case of non-Abelian gauge anomalies. In that case, the consistent anomaly satisfies the Wess-Zumino consistency condition, but fails to be gauge covariant (it involves $dA$ rather than $F$). The covariant anomaly on the other hand, does not satisfy the Wess-Zumino consistency condition, but is fully gauge covariant. The consistent and covariant versions of the anomaly differ by current redefinitions which do not come from local counterterms, but are equivalent so far as anomaly cancellation is concerned. The difference between the covariant and consistent currents is usually referred to as the \textsl{Bardeen-Zumino polynomial}.} We see therefore, that the covariant current in the boundary carries the charge which is injected into it from the bulk.

The Hall viscosity term in (\ref{noReffact3d}) on the other hand is invariantly defined under diffeomorphisms, Lorentz, and \(U(1)\) gauge transformations, and thus does not lead to consistent anomalies in the edge theory. The consistent stress current \(J^a_{cons}\) in the edge is therefore symmetric, and suffers only from the anomaly due to the \(U(1)\) Chern-Simons term
\begin{equation}
D*J^a_{cons} -i_{e^a}T_b\wedge *J^b_{cons} -i_{e^a}F\wedge *J_{cons}= -q^2\frac{n_R-n_L}{4\pi}(i_{\underline{e}^a}A) F \label{consanomaly2}
\end{equation}
\begin{equation}
e^{[a}\wedge *J^{b]}_{cons} =0
\end{equation}
(Recall that since the domain wall is $1+1$-dimensional, the spin current $J^{ab}$ vanishes.) Equation \eqref{consanomaly2} is not gauge covariant. However, it is clear what we must do - we shift to the covariant currents. 

Consider then, the variation of the bulk action under a change in the frame and connection
\begin{equation}
\delta S_{odd,bulk} = \zeta_H\int_M \delta e^A\wedge T_A -\frac{\zeta_H}{2}\int_M \delta \omega_{AB} e^A\wedge e^B +\frac{\zeta_H}{2}\int _{\Sigma}\delta e^A\wedge  e_A \label{bulkvartor}
\end{equation}
We digress momentarily to explain a minor point. In the boundary term above, we should interpret the result in terms of fields defined on the boundary. Generally (as was implied in the discussion of the gauge case above), $p$-forms will pull back to the boundary. In the case of vector-valued forms $e^A$ and $\omega^{AB}$, we also must decompose the pullbacks in representations of the boundary Lorentz group. Generally, we are free to choose independently a co-frame $E^a$ and spin connection $\Omega^{ab}$ in the boundary. These can be identified with the pull-backs of $e^a$ and $\omega^{ab}$ up to a Lorentz transformation. The normal components $e^n$ and $\omega^{na}$ represent extrinsic effects. Conventionally, the pullback of $e^n$ to $\Sigma$ vanishes, which can be achieved by a local bulk Lorentz transformation of the frame.

Returning to Eq \eqref{bulkvartor}, we read off the bulk stress current and spin current 
\begin{equation}
*J^A_{bulk} = \zeta_H \;T^A,\;\;*J^{AB}_{bulk} = -\frac{\zeta_H}{2} e^A\wedge e^B
\end{equation}
while the stress current induced in the edge theory is
\begin{equation}
*j^a = \frac{\zeta_H}{2}\;e^a
\end{equation}
It is easy to check that the bulk currents satisfy the proper (non-anomalous) Ward identities
\begin{equation}
D*J^A_{bulk} - i_{e^A}T_B\wedge *J^B_{bulk} -i_{e^A}R_{BC}\wedge *J^{BC}_{bulk}-i_{e^A}F\wedge *J_{bulk}= 0
\end{equation}  
\begin{equation}
D *J^{AB}_{bulk} -e^{[A}\wedge *J^{B]}_{bulk} =0\nonumber
\end{equation}
But once again, the fluxes into \(\Sigma\) are non-trivial. These are easily computed\footnote{Here, we disregard the normal component $\Delta Q_\Sigma^n$. Since this is related to a bulk diffeomorphism normal to the edge, we expect that it is related to {\it extrinsic} rather than intrinsic edge properties.}
\begin{equation}
\Delta Q^a_{\Sigma} = \zeta_H\int_{\Sigma} T^a \label{frameflux}
\end{equation}
\begin{equation}
\Delta Q^{ab}_{\Sigma} =- \frac{\zeta_H}{2}\int_{\Sigma} e^a\wedge e^b
\end{equation}
%
%
We now write the Ward identities in the edge for the covariant currents \(J^a_{cov}=J^a_{cons}+j^a\) and \(J_{cov}\)
\beqn
d*J_{cov} &=& \sigma_H\;F\\
D*J^a_{cov} -i_{e^a}T_b\wedge *J^b_{cov} -i_{e^a}F\wedge *J_{cov}&=& \zeta_H\; T^a \label{covanomaly2}
\\
e^{[a}\wedge *J^{b]}_{cov} &=& \frac{\zeta_H}{2} e^a\wedge e^b \label{loranomaly2}
\eeqn
Notice that the right-hand sides agree with the ``charge'' (in this case energy-momentum) entering the edge from the bulk. 

Let us now extend the above analysis to include curvature. One finds that the covariant anomalies, when written in terms of torsion and Levi-Civita curvature, become
\beqn
d*J_{cov} &=& \sigma_H\;F\\
D*J^a_{cov} -i_{e^a}T_b\wedge *J^b_{cov} -i_{e^a}F\wedge *J_{cov}&=& \zeta_H\; T^a +\kappa_H\left(e^a\wedge d\lcR-\lcR\; T^a\right)\label{covanomaly2'}
\\
e^{[a}\wedge *J^{b]}_{cov} &=& \frac{1}{2}\left(\zeta_H-\kappa_H\lcR\right) e^a\wedge e^b \label{loranomaly2'}
\eeqn
In the next section, we will show that it is possible to derive these identities from the edge point of view using the Fujikawa method with a suitable choice of regularization. 

Given the above results, it is natural to ask if the torsion terms in the diffeomorphism Ward identity could be removed by the addition of local counterterms. Indeed, a shift of the stress current
\beq
*J^a_{cov} \rightarrow *J^a_{cov}-\frac{1}{2}\left(\zeta_H-\kappa_H\;\lcR\right)e^a
\eeq
would make it symmetric. This shift however {\it does not} come from a local counterterm in the boundary theory, as one can easily see after close  inspection, nor is it an ordinary improvement term.\footnote{An \textsl{improvement} of the stress current is a current redefinition which makes it symmetric, but does not modify it's conservation equation.} In fact, it amounts to shifting the bulk effective action by 
\beq
\Delta S_{odd,bulk}=-\frac{1}{2}\int_M\zeta_H\;e^A\wedge T_A + \frac{1}{2}\int_M\kappa_H\;\lcR e^A\wedge T_A
\eeq
One of our main precepts is that divergences that appear in the bulk are common to all phases. Thus shifting the values of $\zeta_H$, $\kappa_H$, etc. by finite counterterms simultaneously in all phases is allowed, but this does not change the differences in their values between phases. Therefore, we cannot avoid having a torsional response in one of the two phases. It is this invariant information that is encoded in the covariant anomalies of the edge theory, and these are the important physical effects.

It is also interesting to see from equations \eqref{covanomaly2'} and \eqref{loranomaly2'} (also see Eq \ref{parityodd1}) that the viscosity is shifted by a term proportional to the Ricci scalar. So if our space-time manifold is of the form \(\mathbb{R}\times \Sigma\), with \(\Sigma\) being a constant curvature Riemann surface of Euler characteristic \(\chi_E\) and area \(A\), then the viscosity is shifted by an amount proportional to $\frac{\chi_E}{A}$. We will return to this point in our concluding remarks. 

\section{1+1d Anomalies from Fujikawa method} \label{section 8}
We will now derive the covariant Ward identities discussed in the previous section, from the edge point of view. We will use standard methods that produce the covariant anomalies, and the novelty of the calculation is that we will produce the torsional contributions to the anomalies. In so doing, we will encounter divergences associated with the torsional terms. Our context provides these divergences with a clear interpretation, as the ultraviolet cutoff of the edge theory is determined by the mass gap in the bulk, and their presence is linked with bulk transport properties.\footnote{Similarly, in 3+1 dimensions,\cite{chandia1997} torsional contributions to the chiral anomaly have been found, with divergent coefficients, although the interpretation of these divergences was not understood. These divergences can be given a physical interpretation in terms of 4+1d topological insulators.\cite{uslaterwork}}

The chiral fermions localized on a 1+1-d space-time manifold \(\Sigma\) coming from the boundary of the manifold $M$ couple to the co-frame $e^a$ on \(\Sigma\). For simplicity, we will assume that the geometry near \(\Sigma\) is separable, with a co-frame of the form \(e^A=(Ndx,e^a)\). For our purposes, it will also suffice to ignore extrinsic couplings to the chiral fermions because these do not affect the covariant anomaly computations which are of interest here.\footnote{The only extrinsic coupling to the chiral states is through a term in the effective action of the form
\begin{equation}
S_{ext} \simeq \frac{1}{48\pi}\int_{\Sigma} ln(N) \lcR \;vol_{\Sigma} \label{ext}
\end{equation}
where \(\lcR\) is the Ricci scalar on $\Sigma$. This term ensures that while the edge theory is anomalous under a Nieh-Yan-Weyl transformation of the frame on \(\Sigma\), there is no anomaly due to a Nieh-Yan-Weyl transformation of the bulk frame.}

Let us quickly review the Fujikawa method for computing covariant anomalies. Our discussion here mainly follows \cite{AGW, AGPM}. The main point of the Fujikawa method is that the variation of the fermion measure under symmetry transformations leads to anomalous Ward identities. For a Dirac fermion \(\Psi\), one defines the measure as follows: expand \(\Psi\) and \(\bar\Psi\) in terms of eigenfunctions \(\Phi_m\) of a self adjoint operator, conventionally chosen to be the Dirac operator
\begin{equation}
\slashed \msD \Phi_m = \lambda_m \Phi_m
\end{equation}
\begin{equation}
\Psi = \sum_m a_m \Phi_m,\;\;\bar\Psi = \sum_n b_n \bar\Phi_n
\end{equation}
and define the measure as \([d\Psi d\overline{\Psi}]= \prod_{m,n} db_mda_n\). However for a left-chiral fermion \(\psi\), the operator \(\slashed \msD_L =\slashed \msD \frac{1}{2}(1-\gamma^5)\) is not self-adjoint. Thus, \(\psi\) must be expanded in terms of eigenfunctions \(\phi_m\) of \(\slashed \msD_L^{\dagger}\slashed \msD_L\) and \(\bar\psi\) must be expanded in terms of eigenfunctions \(\chi_n\) of \(\slashed \msD_L\slashed \msD_L^{\dagger}\). Under a symmetry transformation \(T:\psi \rightarrow \psi'=\psi+\delta_T\psi\), the measure could transform in general
\beq
[d\psi'd\overline{\psi}'] = e^{-i\int \mathcal{A}_T}[d\psi d\overline{\psi}]
\eeq 
When this happens, the classical Ward identity gets modified by the anomalous correction \(\mathcal{A}_T\). Let us now study this in detail for diffeomorphisms, local Lorentz transformations, and Nieh-Yan-Weyl transformations.
\subsection{Diffeomorphisms}
Recall from \eqref{covdiff} that under a covariant diffeomorphism generated by a vector field $\underline{\xi}$, we have
\begin{equation}
\delta \psi = \nabla_{\underline{\xi}}\psi
\end{equation}
For simplicity, we restrict ourselves to volume preserving diffeomorphisms. Then to linear order in \(\underline{\xi}\), it is easy to check that the measure transforms as
\begin{equation}
[d\psi'd\overline{\psi}']= \exp\left(-\sum_m\int vol_{\Sigma}\bar\phi_m\xi^i\nabla_i\phi_m+\sum_n\int vol_{\Sigma}\bar\chi_n\xi^i\nabla_i\chi_n\right)[d\psi d\overline{\psi}]
\end{equation}
Thus the corresponding Ward identity \eqref{ClassicalDiffWI} gets modified to
\begin{equation}
i\int_{\Sigma}\xi^a(D*J_a- i_{e_a}T^b\wedge *J_b-i_{e_a}F\wedge *J) =\sum_m\int_{\Sigma} vol_{\Sigma}\left(\bar\phi_m\xi^a\nabla_a\phi_m-\bar\chi_n\xi^a\nabla_a\chi_n\right)=-\mathrm{Tr}\;\gamma^5\xi^a\nabla_a \label{QuantumDiffWI}
\end{equation}
Clearly, the trace is ill-defined by itself, and needs to be regulated. Customarily, it is regulated using the heat-kernel regularization in Euclidean space
\begin{equation}
-\mathrm{Tr}\;\gamma^5 \xi^a\nabla_a e^{\slashed{\msD}^2/ \Lambda^2}
\end{equation}
where $\Lambda$ the ultraviolet cutoff is taken off to infinity. However for the edge theory we consider, the ultraviolet cutoff \(\Lambda\) is of order \(m\), the bulk mass gap, since the spectrum of \emph{localized edge modes} of the bulk Dirac operator only exists for energies $E<\vert m\vert.$ This issue is irrelevant in the torsionless case, because the leading terms in the anomaly are finite and cutoff independent. However, in the presence of torsion we find a quadratic divergence in the anomaly if regulated naively. Moreover, the divergent term cannot be removed by a local counterterm. 

The guiding principle in choosing the appropriate regularization must then be bulk-boundary matching --- the non-conservation of ``charge'' (in the present case, energy-momentum) as manifested in the covariant anomaly must match the influx of charge due to the parity violating terms in the bulk action. Fortunately, a minor generalization of the results in \cite{AGPM} readily implies that (in a separable geometry) the flux of the bulk stress current in a given phase into the edge is given by
\begin{equation}
-\sum_{i=0}^N\frac{1}{2}C_i\mathrm{sign}(M_i)\mathrm{Tr}\;\gamma^5\xi^a\nabla_a\frac{1}{(1-\slashed{\msD}^2/M_i^2)^{1/2}} 
\end{equation}
where \(M_i\) are the masses of the bulk fermions, including Pauli-Villars regulators. Notice that this is exactly the trace that was obtained in the Fujikawa formalism, albeit in a regulated form. Therefore, it is clear that we must regulate the Ward identity \eqref{QuantumDiffWI} as
\begin{equation}
i\int_{\Sigma}\xi^a(D*J_a- i_{e_a}T^b\wedge *J_b-i_{e_a}F\wedge *J) = -\Delta\sum_{i=0}^N\frac{1}{2}C_i\mathrm{sign}(M_i)\mathrm{Tr}\;\gamma^5\xi^a\nabla_a\frac{1}{(1-\slashed{\msD}^2/M_i^2)^{1/2}}\label{diff}
\end{equation}
with \(M_0 = m\) the mass gap in the bulk, and \(M_i\) for \(i=1,\cdots N\) being the masses of the Pauli-Villars regulator fermions in the bulk. The symbol \(\Delta\) indicates that the anomaly is the difference between the flux from the non-trivial phase and the flux from the trivial phase.  

In order to compute the trace (in Euclidean space), we can rewrite it as
\beq
-\Delta\sum_{i=0}^N\frac{1}{2\Gamma(\frac12)}C_iM_i\int_{\epsilon}^{\infty}dt\;t^{-1/2}\mathrm{Tr}\;\gamma^5\nabla_ae^{-t(-\slashed{\msD}^2+M_i^2)} \label{difftrace}
\eeq
The asymptotic expansion corresponding to this trace in 2 dimensions is given by (see Appendix \ref{sec: AE})
\beq
\mathrm{Tr}\;\gamma^5\xi^a\nabla_ae^{t\slashed{\msD}^2}\simeq -i\int_{\Sigma} \xi^a\left(\frac{1}{4\pi t}T_a+\frac{1}{48\pi}e_a\wedge d\lcR - \frac{1}{48\pi}\lcR\;T_a+\cdots\right)
\eeq
where the ellipsis denote terms higher order in \(t\) (which are unimportant here as they will give higher order terms suppressed by inverse powers of the cutoff). The integral over \(t\) in \eqref{difftrace} diverges as $\epsilon\rightarrow 0$ for the first term above, but the divergence is cancelled by the condition $\sum_{i=0}^N C_iM_i=0$ on the regulator masses. Using the expressions for regulated bulk coefficients (\ref{regcoefffull1}, \ref{regcoefffull2}, \ref{regcoefffull3}), we get the Ward identity
\beq
(D*J_a- i_{e_a}T^b\wedge *J_b-i_{e_a}F\wedge *J) = \zeta_H\;T_a + \kappa_H\;(e_a\wedge d\lcR-\;\lcR\;T_a)
\eeq
where \(\zeta_H\) and \(\kappa_H\) are the regulated coefficients in the non-trivial phase. Note that this is exactly what we found in our analysis in the previous section (see \eqref{covanomaly2'}). 

\subsection{Local Lorentz transformations}
The change in the measure corresponding to Lorentz transformations \(\delta \psi= \frac{1}{4}\theta_{ab}\gamma^{ab}\psi\) is given by
\beq
[d\psi'd\overline{\psi}'] = \exp\left(-\frac{1}{4}\mathrm{Tr}\;\theta_{ab}\gamma^5\gamma^{ab}\right)[d\psi d\overline{\psi}]
\eeq
Following the discussion of the diffeomorphism anomaly in the previous section, we regulate the Ward identity as
\begin{equation}
i\int_{\Sigma} \theta_{ab}\;e^{[a}\wedge *J^{b]} = -\frac{1}{4}\Delta\sum_{i=0}^N\frac{1}{2}C_i\mathrm{sign}(M_i)\mathrm{Tr}\;\theta_{ab}\gamma^5\gamma^{ab}\frac{1}{(1-\slashed{\msD}^2/M_i^2)^{1/2}}\label{lor}
\end{equation}
Using the asymptotic expansion in 2 dimensions
\beq
\mathrm{Tr}\;\gamma^5\gamma^{ab}e^{t \slashed{\msD}^2} \simeq \int_{\Sigma} -i\epsilon^{ab}\left(\frac{1}{2\pi t}-\frac{1}{24\pi}\lcR+\cdots\right)vol_{\Sigma}
\eeq
we find \eqref{loranomaly2'}
\begin{equation}
e^{[a}\wedge *J^{b]}_{cov} = \frac{1}{2}\left(\zeta_H-\kappa_H\lcR\right) e^a\wedge e^b 
\end{equation}
\subsection{Nieh-Yan-Weyl transformations}
Under a Nieh-Yan-Weyl transformation \(\delta\psi = -\frac{1}{2}\sigma \psi\), the measure transforms as\footnote{It is important to note that both chiralities contribute with the same sign to the Nieh-Yan-Weyl anomaly, which is therefore proportional to \((n_L+n_R)\). On the other hand, the anomalies corresponding to diffeomorphisms and local Lorentz transformations come with \((n_L-n_R)\).} 
\beq
[d\psi'd\overline{\psi}'] = \exp\left(\frac{1}{2}\mathrm{Tr}\;\sigma\right)[d\psi d\overline{\psi}']
\eeq
The situation with Nieh-Yan-Weyl transformations is slightly different - the trace here cannot be interpreted in terms of the flux from bulk parity violating terms. However, there is no problem with regulating the anomalous Ward identity by the traditional heat kernel method 
\beq
i\int \sigma\;e_a\wedge *J^a = \frac{1}{2}\mathrm{Tr}\;\sigma e^{\slashed{\msD}^2/\Lambda^2}
\eeq
This is because the quadratically divergent terms in this case can be removed by an appropriate local counterterm. We will therefore only be interested in the finite and universal piece in the trace. Using the asymptotic expansion
\beq
\mathrm{Tr}\;e^{t \slashed{\msD}^2}\simeq  \int_{\Sigma} \left(\frac{1}{2\pi t}-\frac{1}{24\pi}\lcR+\cdots\right)\;vol_{\Sigma}
\eeq
and shifting the stress current by a local counterterm \(J^a \rightarrow J^a - \frac{\Lambda^2}{4\pi} e^a\), we get
\beq
e_a\wedge *J^a = -\frac{1}{48\pi}\; \lcR\;vol_{\Sigma}
\eeq
Indeed, this agrees with the well-known result $ {T^{\mu}}_{\mu} = -\frac{(c_L+c_R)}{48\pi}\lcR$, because for a left-handed Weyl fermion \((c_L,c_R)=(1,0)\).

\section{Spectral Flow}\label{sec:spectralflow}

Given the close analogy between the Hall conductivity and Hall viscosity that we have discussed above, it seems reasonable that there ought to be physical adiabatic processes that induce spectral flow in the interface in each case. The case of the Hall conductivity is well-known, and we briefly review it here.

\subsection{Spectral Flow and Hall Conductivity}
We consider a gauge field on a spatial cylinder of length $L$ in the $x$-direction and radius $R$ in the $y$-direction
\beq
A=E_ytdy
\eeq
where $E_y$ is a constant. This is equivalent to
\beqn
F&=& E_y dt\wedge dy\\
*_3F&=&-E_y dx
\eeqn
Thus we have a constant electric field in the $y$-direction which we imagine resulting from the threading of electromagnetic flux along 
the cylinder. We can parameterize $E_y=-\frac{h}{2\pi qR T}$ where $q$ is the charge and $T$ is the time it takes to thread one flux quantum into the hole of the cylinder. 

Given the effective action for the bulk charge response $S_{eff}[A]=\int_M\left(\frac12\sigma_H A\wedge dA\right)$, where $\sigma_H=-\frac{q^2}{h}$ and $A$ is the electro-magnetic gauge field, the expectation value for the charge current is given by $J=\sigma_H *_3F$. Thus we find the bulk current response to the electric field is
\beq
J=\frac{q^2}{2\pi \hbar}\frac{\hbar}{qRT}dx=\frac{q}{2\pi RT}dx
\eeq
This means there is a constant current density in the $x$-direction, and over the time $T$ we build up charge 
\beq
\Delta Q=\int_0^T dt \int_0^{2\pi R} dy J_x=q. 
\eeq

From the point of view of the intrinsic boundary theory, which consists of chiral fermions of one chirality or the other, this increase in charge is an anomalous process. In $1+1$, the chiral symmetry is anomalous $d*_2 J_5=\frac{q}{2\pi\hbar} F_{(2)}$, where $F_{(2)}$ is the gauge curvature in $1+1$. In the present case, this is the pull-back of the bulk field, which is just $F_{(2)}=E_y dt\wedge dy$. So the axial charge changes in this process by $\Delta Q_5=\int_{\Sigma_x}d*_2J_5=-1$. This change occurs as a chiral fermion is pumped from one edge of the cylinder to the other through the bulk Chern-Simons response.

\begin{figure}[t]
\begin{center}
\includegraphics[width=6.45in]{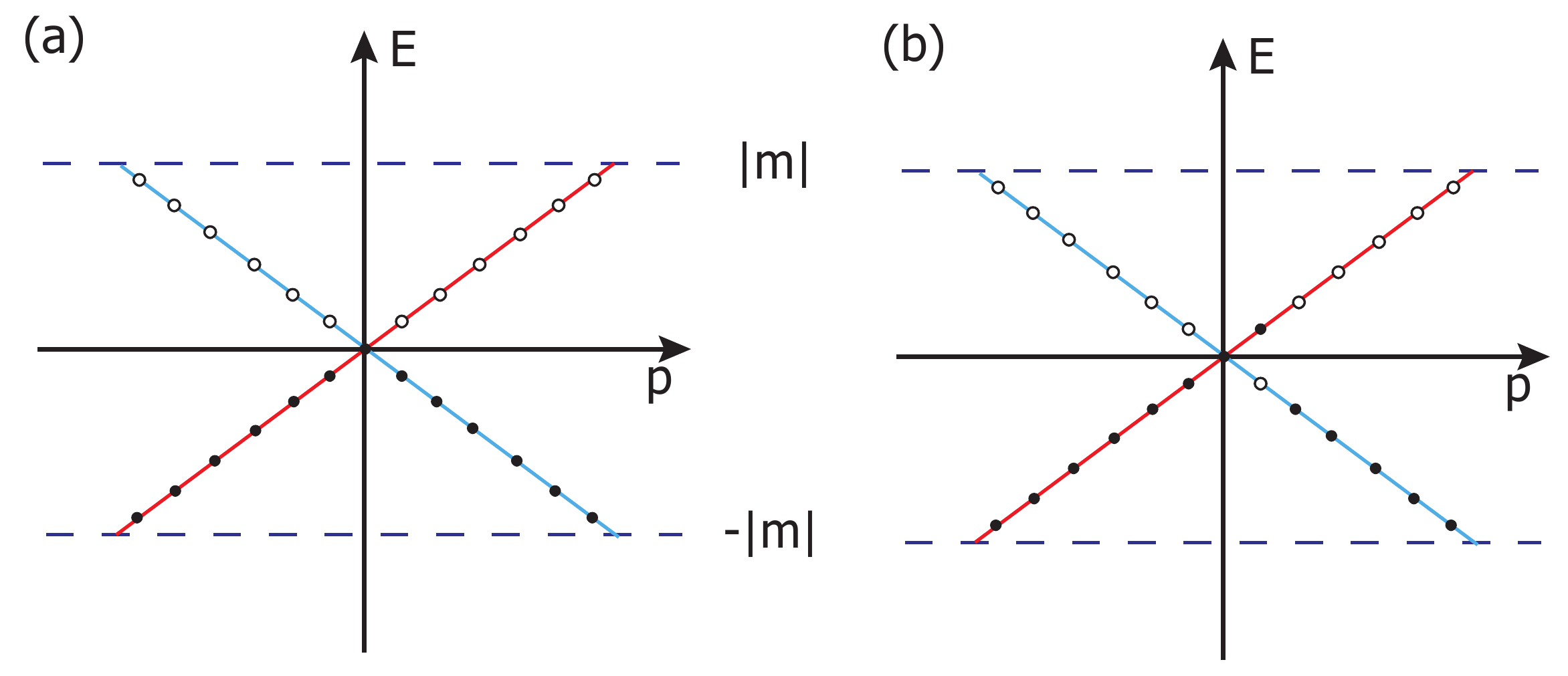}
\caption{(a) Energy spectrum from Eq.\ (\ref{eq:chiralHgauge}) at time $t=0.$ Right/left handed fermion spectra are represented by  positively/negatively sloped lines. The filled/empty circles represent occupied/unoccupied states. (b) Energy spectrum at time $t=T$ where one flux quantum has been threaded through the spatial ring. The spectrum returns to itself but the state occupation changes. One electron has been added to the right movers, and one has been removed from the left movers.} \label{fig:flowgauge}
\end{center}
\end{figure}

We can get a simpler pictorial understanding of the anomaly by considering the energy spectrum of the chiral fermions. The Hamiltonians for the left- and right-handed chiral fermions are 
\begin{equation}
H_{R}=v(p-qA)\;\;\; \ \ \  H_{L}=-v(p-qA)\label{eq:chiralHgauge}
\end{equation}\noindent where the vector potential is $A=\frac{\hbar t}{qRT}.$ Substituting this form into Eq. \ (\ref{eq:chiralHgauge}) we find
\begin{equation}
H_{R}=\frac{\hbar v}{R}\left(n-\frac{t}{T}\right)\;\;\; \ \ \ \  H_{L}=-\frac{\hbar v}{R}\left(n-\frac{t}{T}\right)
\end{equation}\noindent where $n$ is an integer labeling the discrete momentum modes $p=\frac{2\pi n\hbar}{2\pi R}.$ Assuming that $T$ is very large so that the spectrum changes adiabatically, we find that the spectrum \emph{flows} as time increases. At a time $t=T$, or in fact at any multiple of $T,$ the spectrum returns to its initial configuration, yet the system as a whole has changed because the state occupation changes. When $t=rT$ for integer $r$ there have been $r$ flux-quanta threaded into the circle on which the chiral fermions live. For each flux quanta threaded an electron is transferred from the left movers to the right movers as illustrated in Fig. \ref{fig:flowgauge}. Thus we reproduce our calculations from above by observing the transfer of electrons during the spectral flow process.  

\subsection{Spectral Flow and Hall Viscosity}

\begin{figure}[t]
\begin{center}
\includegraphics[width=6.45in]{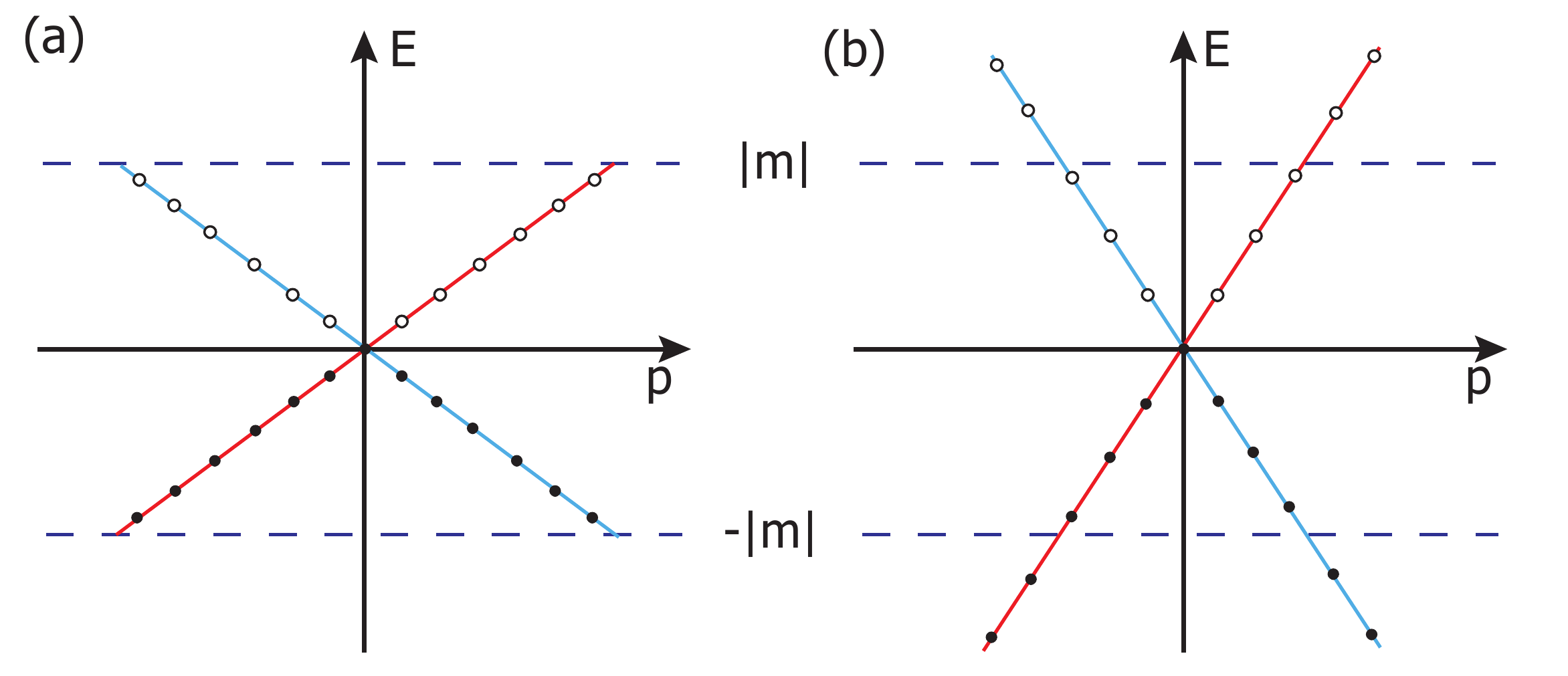}
\caption{(a)Energy spectra for left (blue) and right (red) handed chiral fermions from Eq.\ \ref{eq:torsionHam} at $t=0$  (b) Energy spectra for $t>0$ assuming $b_L=b_R=b<0$ which gives an increase to the velocity for both branches of chiral fermions. Note that when compared to the $t=0$ case there are states that were occupied chiral fermion states that have been pushed past the cut-off scale and the states at $p=0$ are unchanged. During this process no states cross $E=0$ and there is not a conventional notion of spectral flow at low-energy. Both figures have the same momentum discretization spacing, but different velocities which leads to a different number of states within the cut-off window. } \label{fig:flow}
\end{center}
\end{figure}
Now we would like to understand the momentum transport due to the Hall viscosity using a spectral-flow type argument similar to the case of charge transport. On the spatial cylinder of length \(L\) in the \(x\)-direction and radius \(R\) in the \(y\)-direction, consider the co-frame 
\beq
e^{0}=dt,\; e^1=dx,\; e^{2}=(1+h(t))dy
\eeq
where we will parameterize $h(t)=\frac{bt}{2\pi R T}$ where $T$ is a very large time-scale so that the change is adiabatic, and $b$ has units of length. For simplicity, we will choose the connection \(\omega_{AB}=0\), for which the given co-frame is torsional. This configuration represents the threading of torsion flux \(T^2= \frac{b}{2\pi RT}dt\wedge dy\), \emph{i.e.} a dislocation into the ring with a time-dependent Burgers' vector tangent to the ring with length $bt/T$ at time $t.$ To calculate the bulk energy-momentum flow we must introduce a \textsl{covariant Killing vector field}\footnote{We call a vector field \(\xi\) \textsl{covariantly Killing} if the co-frame is preserved under a covariant diffeomorphism along \(\xi\), i.e. if \(D\xi^a+i_{\xi}T^a=0\).} \(\xi=\xi_a\;\underline{e}^a = \partial_y\). From our previous discussion of Hall viscosity response, the energy-momentum flux along \(\xi\) through a constant \(x\) slice \(\Sigma_x\) is given by
\beq
\int_{\Sigma_x}\xi_a*J^a =\int_{\Sigma_x}\xi_aT^a= \zeta_H\int_{\Sigma_x} \left(1+\frac{bt}{2\pi RT}\right)\frac{b}{2\pi RT} dt\wedge dy 
\eeq 
This leads to a transfer of momentum from one edge to the other through the bulk of the cylinder. From the point of view of the edge chiral fermions localized at \(x=0\) and \(x=L\), this is an anomalous process. For instance at \(x=0\), define the chiral momentum \(P_L(t) = \int_{\gamma_t} \xi_a*J^a\), where \(\gamma_t\) is the spatial circle at time \(t\) and \(J^a\) the intrinsic 1+1d stress current on \(\Sigma_{x=0}\). Then the anomalous conservation law \eqref{covanomaly2} becomes (in the absence of \(U(1)\) gauge fields)
\beq
\frac{dP_L}{dt} = \zeta_H \int_0^{2\pi R} \left(1+\frac{bt}{2\pi RT}\right)\frac{b}{2\pi RT}dy = \zeta_H\left(1+\frac{bt}{2\pi RT}\right)\frac{b}{T} \label{momchange1}
\eeq

Let us now understand the anomalous momentum transfer in terms of the spectra of the left and right handed chiral fermions localized at \(x=0\) and \(x=L\) respectively, with co-frame fields parameterized  by $b_L$ and $b_R$ with the Hamiltonians
\begin{eqnarray}
H_{R}&=&\frac{\hbar v k}{1+\frac{b_{R}t}{2\pi RT}}\\
H_{L}&=&-\frac{\hbar v k}{1+\frac{b_{L}t}{2\pi RT}}\label{eq:torsionHam}
\end{eqnarray}\noindent where $v$ is the chiral fermion velocity and we have assumed the Hamiltonian is acting on translationally invariant plane-wave states. We show the energy spectra at two different times in Fig. \ref{fig:flow}(a) and (b). In the figure we have indicated a high-energy cut-off governed by the scale $|m|.$ This scale represents the energy at which the edge states of a topological insulator merge with bulk states and are no longer localized on the edge. In fact, for such topological insulators like the lattice Dirac model, the cut-off is \emph{exactly} the insulating mass scale $\vert m\vert.$ We have assumed that the energy states are filled up to energy $E=0$ as indicated by the filled circles in Fig. \ref{fig:flow}. The range of momenta that is occupied by right (left) movers is between $p\in\left[-\frac{m}{v}\left(1+\frac{b_L t}{2\pi RT}\right),0\right]$ $(p\in\left[0, \frac{m}{v}\left(1+\frac{b_R t}{2\pi RT}\right)\right])$ . The total momenta of the right and movers at time $t$ is
\begin{eqnarray}
P^{(tot)}_{R}&=&\frac{2\pi R}{2\pi\hbar}\int_{-\frac{m}{v}\left(1+\frac{b_R t}{2\pi RT}\right)}^{0} p dp=-\frac{R}{\hbar}\left[\frac{m}{v}\left(1+\frac{b_R t}{2\pi RT}\right)\right]^2\nonumber\\
P^{(tot)}_{L}&=&\frac{2\pi R}{2\pi\hbar}\int_{0}^{\frac{m}{v}\left(1+\frac{b_L t}{2\pi RT}\right)} p dp=\frac{R}{\hbar}\left[\frac{m}{v}\left(1+\frac{b_L t}{2\pi RT}\right)\right]^2
\end{eqnarray}

As we have seen, the Hall viscosity is related to a stress-energy response and thus to the rate of change of momentum. We find
\begin{eqnarray}
\dot{P}^{(tot)}_{R}&=&-\left(\frac{m}{\hbar v}\right)^2\left(1+\frac{b_L t}{2\pi R T}\right)\frac{\hbar b_L}{2\pi T}\nonumber\\
\dot{P}^{(tot)}_{L}&=&\left(\frac{m}{\hbar v}\right)^2\left(1+\frac{b_R t}{2\pi R T}\right)\frac{\hbar b_R}{2\pi T}\label{eq:momchange2}
\end{eqnarray}\noindent  We see that if we choose $b_{L}\neq b_{R}$ then momentum is not conserved at all if we only consider the edge states and take into account transfers between the edges. Momentum of course is still conserved globally because the excess/deficient amount of momentum gets trapped on some extra torsional flux that will appear in the gapped bulk region away from the edges when $b_L\neq b_R.$ For now we will fix $b_{L}=b_{R}=b$ to avoid this extra complication.

\noindent Comparing equations \eqref{momchange1} and \eqref{eq:momchange2}, we see that the bulk and boundary momentum transport only matches for 
\begin{equation}
\zeta_{H}=\frac{\hbar}{2\pi}\left(\frac{m}{\hbar v}\right)^2
\end{equation}\noindent which is the same result we calculated earlier for the regulated Hall viscosity albeit with all the factors of $\hbar$ and velocity (speed of light) added back in.

While at first glance it appears strange that the viscosity depends on the mass, we can clearly see the reason why this dependence is necessary by examining Fig.\ \ref{fig:flow}. The effect of threading a torsional flux (\emph{i.e.} threading a dislocation) into the loop on which the chiral fermions propagate can be interpreted in one of two ways. Our choice of torsional flux (\emph{i.e.} our specific choice of frame) means that if we travel around the loop at time $t=T$ we enclose a Burgers' vector that is tangent to the ring of length $b.$ Depending on the sign of $b$ this implies that the ring looks either shorter or longer than its original length at $t=0.$ From this perspective we would think of chiral fermions with a fixed velocity but propagating on a ring with a time-dependent length (which will re-discretize the momentum modes as a function of time). The other interpretation is that the length stays fixed at $2\pi R$ but the chiral fermions are either traveling faster or slower depending on the sign of $b.$ This is the interpretation represented in Fig.\ \ref{fig:flow} where the velocity of the chiral fermions has increased at a later time but the momenta retain the original quantization scale. Thus we see that coupling to the $U(1)$ electromagnetic field causes a \emph{translation} in the spectrum, but the coupling to torsion causes a \emph{scaling} of the spectrum. As a function of time the two chiral branches rotate in opposite directions around the fixed point where $p=0.$ This is because $p=0$ does not feel any effects of torsion since it is \emph{uncharged} as far as torsion is concerned.  So the torsional response is given by spectral scaling/rotation instead of spectral flow/translation. In terms of the discussion we used in the introduction this occurs because each momentum mode carries a different charge under torsion, while they all carry the same $U(1)$ gauge charge. In fact, the state at $p=0$ does not even see the torsional flux and is unmodified since it carries zero torsional charge. 

\subsection{Comments on Spectral Flow}
\begin{figure}[h]
\begin{center}
\includegraphics[width=6.45in]{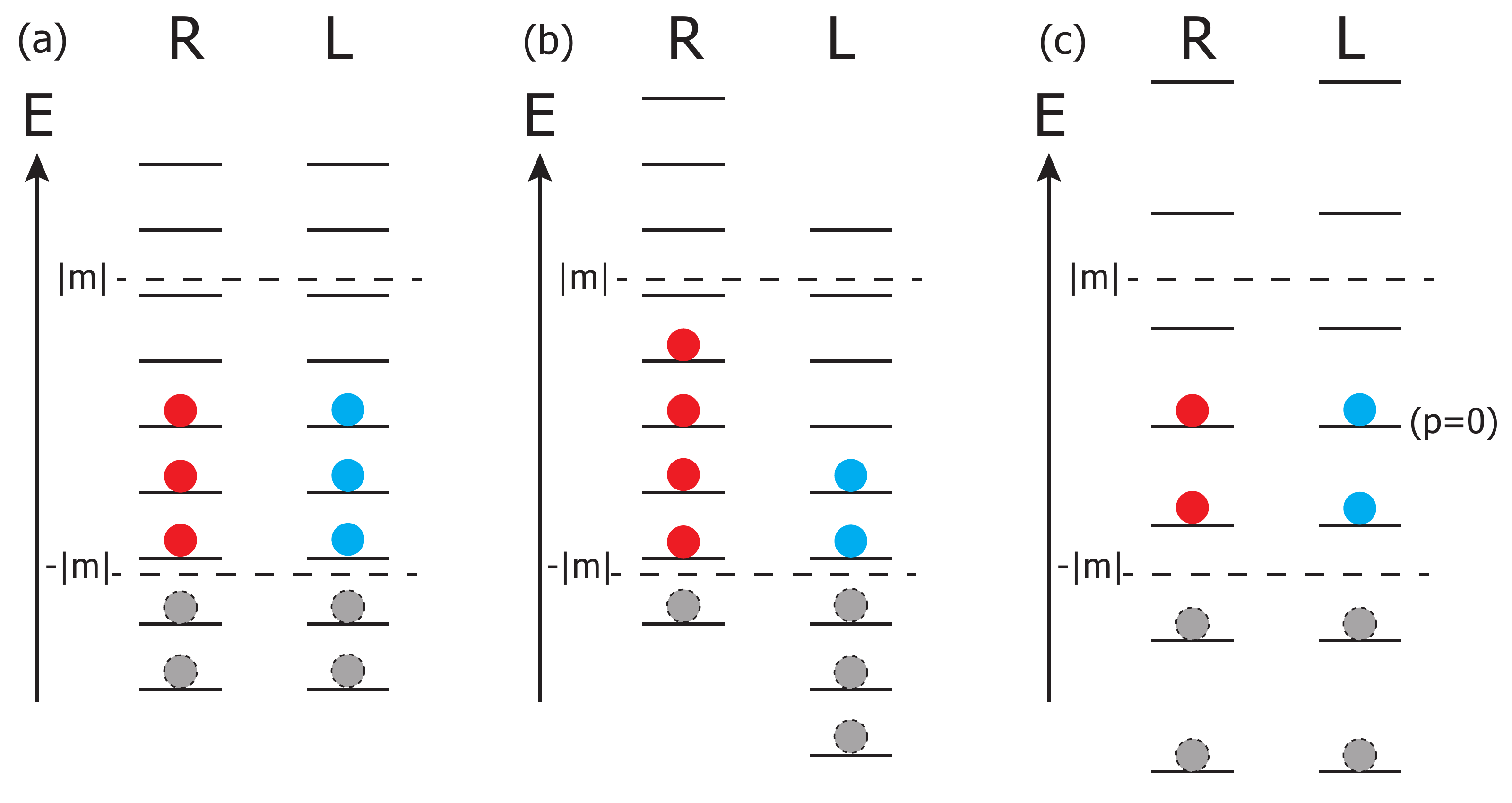}
\caption{Schematic illustration of the energy spectra for the case of (a) unmodified chiral fermions  (b) insertion of an electromagnetic flux quantum (c) insertion of torsional flux. The red/blue circles represent occupied Right/Left handed chiral fermion states. The gray-dashed circles represent occupied states which are not chiral since they are beyond the energy cutoff at $E=\pm \vert m\vert$ (represented by the dashed horizontal lines). Empty lines represent unoccupied modes. Note that in (b) none of the states are fixed and they all shift by the same amount whereas in (c) the states at $p=0$ are fixed and the states shift by amounts proportional to their momenta. } \label{fig:stretch}
\end{center}
\end{figure}
In this section we would like to discuss some comparisons between the anomalous electromagnetic and torsional responses in terms of the behavior of the edge state spectra. Anomalies connect low-energy physics with high-energy physics, and thus understanding both regimes is important for discerning the physics. For chiral fermions arising as the boundary theory of a higher dimensional massive Dirac model, both regimes can be understood. The low-energy physics is that of linearly dispersing chiral fermions that propagate along the 1+1-d boundary; this is the theory we have been carefully studying. In the high-energy regime there is an energy cutoff when $E=\pm\vert m\vert$ at which the chiral branch of states, which was localized on the boundary of the system, spreads into the higher-dimensional bulk and couples with the boundary states on the opposite edge. At this energy, and beyond, the fermions are no longer chiral, but in fact form Dirac fermions through their coupling with the opposite chirality edge states on the other boundary. Thus, states which lie outside of $E=\pm \vert m\vert$ are not chiral, and not localized on an edge. 

In Fig.\ \ref{fig:stretch} we compare the behavior of the spectra for both anomalous processes. Fig.\ \ref{fig:stretch}a shows the unmodified spectrum at $t=0$ for left and right handed branches while Fig.\ \ref{fig:stretch}b,c show the modified spectra at a later time for the electromagnetic and torsional cases respectively. The red/blue colored circles represent occupied states, while the gray (dashed) circles represent occupied states beyond the cut-off which are not chiral. Empty lines represent unoccupied states. The modified spectrum for the electromagnetic case is simply a shift compared with the original spectrum, and is easy to understand since \emph{every} state, no what the energy or momentum, shifts the \emph{exact} same amount. The reason the shift is uniform over the entire spectrum is that all of the states carry the same charge $q$ under the electromagnetic gauge field, and thus all couple minimally with the identical coupling constant. During the process of threading one flux quantum we see that one unoccupied state from the R branch passes beyond the cut-off and becomes non-chiral, while a formerly non-chiral occupied state in the R branch enters the edge state range and attains a right-handed chirality. That is to say there is an additional chiral fermion in the R branch after the process. The opposite phenomenon occurs for the L branch. Since the momenta of all states shift the same way, the notion of whether an anomalous charge appears or does not appear is independent of the width of the cut-off (though it does matter where the cutoff is centered around, which is to say the value of the chemical potential compared to the cut-off can be important). Thus we can easily imagining sending $\vert m\vert\to \infty$ and we would still draw the same conclusion that one charge is produced for each inserted flux quantum. This even makes some sense (although it is imprecise) if we start with the cut-off at infinity since we know the states at the highest energies are shifted by the same amount as the states at the lowest energies since they all have the same electromagnetic charge. 

The torsional case is more complicated to understand. In this case the spectrum is stretched away from the states at $p=0.$ Unlike the electromagnetic case, where none of the states remain fixed when flux is inserted, under torsional flux insertion the states at $p=0$ do not change. Since each state is charged differently under torsion, \emph{i.e.} the momenta of all the single-particle states are different, each state translates by a different amount. Of course, a momentum dependent translation that depends linearly on $p$ is nothing but a scaling of the momentum, and thus a scaling of the energy since $E_{R/L}=\pm vp.$ We immediately run into a problem if we do not have a cut-off because the states at higher energies have higher charges and thus it is not clear how to interpret spectral changes if we allow $p\to\infty.$ For any well-defined system, such as a real material, or a properly regularized theory, this is not a problem because the momentum range over which chiral fermions exist will always be finite. 

As shown in Fig. \ref{fig:stretch}c, when the torsional flux is inserted some states are scaled beyond the cut-off and are no longer localized, chiral modes. If we pushed the cut-off further out in energy there will be more and more states that get scaled beyond the cut-off per unit time since states in a larger range of energy near $\pm \vert m\vert$ will have momenta that are large enough to scale them beyond the cut-off. We see that while the same states stay occupied, and interestingly, no states cross $E=0,$ the amount of states which are actually localized on the edge is reduced. Since none of the states flow through $E=0,$ a connection to an index theorem, if one exists, must arise from a new mechanism. Note that if we considered the opposite sign of the Burgers' vector the states would be scaled downward in energy so that the number of states within the cutoff would be denser.  Thus, the momentum transport does not come from a change in occupation of the states due to a spectrum shift, but instead from a decrease/increase in the density of chiral modes that lie within the cut-off. The electromagnetic modification to the spectrum thus acts like a rigid, incompressible flow of levels  while the torsional modification allows for changes in the density of levels around some fixed point; in both cases the cut-off energy acts as a source/sink of levels. The change in the momentum of a chiral branch depends crucially on how many levels pass through the cutoff and what momenta those levels carry. Since both the number of states, and the amount of momenta the states carry, is linearly proportional to the cut-off energy $\vert m\vert$ we expect that the viscosity, which represents the rate of momentum change, would be proportional to $\vert m\vert^2$ as we have already found.

\section{Conclusions and Remarks}

In this paper, we have presented detailed calculations of physical quantities in Chern insulators, particularly in 2+1 dimensions. In this model, the effective action (generating functional) takes a Lorentz-invariant form. The leading parity-violating effect, the Hall viscosity, is encoded in this effective action through a term involving torsion. The interpretation of this term has been hampered by the appearance of power divergences. This should be compared with the (integer) quantum Hall systems, whereby similar computations give finite answers directly. This occurs simply because there is an effective cutoff set by the magnetic length, and the computations do not encounter divergences because there are a finite number of states (on the spatial torus, say). Nevertheless, careful management of the divergences that appear in the Chern insulator lead to a consistent picture in which the difference in Hall viscosities of distinct topological phases is independent of the cutoff and set by the mass gap. As one might well expect, this difference is also encoded in the structure of anomalies on an edge between such topological phases. Indeed, we have shown that the covariant diffeomorphism anomaly of a 1+1 chiral fermion includes a torsional term that fully accounts for the jump in Hall viscosity across the edge, just as the chiral anomaly accounts for the jump in Hall conductivity.

The appearance of a chiral gravity action with negative cosmological constant in the leading terms of the effective action is intriguing. Certainly this is consistent with the chiral nature of the edge theory, encoding $c_L-c_R=1$. Is there some relation here with dynamical/holographic gravitational systems? In our context of encoding transport coefficients, the answer is surely negative -- there is no role played by gravitational field equations or any particular solution thereof. However, given that dislocation/disclinations are analogous to vortices, effective dynamics may emerge as the result of a condensation mechanism \cite{Zaanen}.
 We will report on this idea elsewhere\cite{uslaterwork}.

One of the features that we have encountered above is the presence of an anti-symmetric part for the stress-energy tensor. We reiterate here that its presence is not forbidden by principle but is usually not considered simply by assumption of the nature of the matter being discussed. However, in any medium in which the microscopic degrees of freedom have local spin degrees of freedom, we can expect that such effects may be manifested. This would apply both to materials as well as presumably to hydrodynamic fluids. 

Finally, let us emphasize an important feature of the calculation of the effective action. In section \ref{sec: chiral gravity}, 
we noted that the effective action organizes itself in terms of the connection $\omega^{(-\torcplg)}$. When written out in terms of torsion and the Levi-Civita connection, the parity odd effective action to linear order in torsion is given by
\beqn
S_{odd}[e,\omega,A]= \frac{\sigma_H}{2}CS[A]+\frac{\kappa_H}{2}CS[\lcw]+\frac{\zeta_H}{2} \int  e^a\wedge T_a- \frac{\kappa_H}{2}\int  \lcR\; e^a\wedge T_a+\cdots \label{poddcon}
\eeqn
where the coefficients \(\kappa_H\) and \(\zeta_H\) have been evaluated in section \ref{sec: chiral gravity}. On a space-time of the form $\mathbb{R}\times \Sigma$, with $\Sigma$ a constant curvature Riemann surface of Euler characteristic $\chi_\Sigma$ and area $A$, \eqref{poddcon} becomes 
\beq\label{Sodd}
S_{odd}[e,\omega,A]= \frac{\sigma_H}{2}CS[A]+\frac{\kappa_H}{2}CS[\lcw]+\frac{1}{2}\left(\zeta_H-\frac{4\pi\kappa_H\chi_{\Sigma}}{A}\right) \int  e^a\wedge T_a +\cdots 
\eeq
We thus find a \emph{shift} in the effective Hall viscosity $\mbox{\boldmath$\zeta_{H}$}$ relative to it's flat space value 
\beq
\mbox{\boldmath$\zeta_{H}$} = \zeta_H -\frac{4\pi\kappa_H\chi_{\Sigma}}{A}
\eeq
This effect is reminiscent of the Wen-Zee shift of the number density in a quantum Hall fluid in the presence of curvature. Let us define the \emph{spin density} \(\mathfrak{s}\) of the Chern insulator as
\beq
\mathfrak{s} = \frac{1}{A}\int_{\Sigma} *J^{12}
\eeq
where \(J^{12}\) is the spatial component of the spin current \(J^{ab}\) (see Eq \eqref{spin current}). This may be computed from the action\footnote{In particular, $J^{ab}$ is obtained by varying with respect to $\omega_{ab}$, holding $e^a$ fixed.} (\ref{Sodd}), and we see that  then the spin density is shifted similarly, and in fact the effective Hall viscosity satisfies 
\beq
\mbox{\boldmath$\zeta_{H}$} = -\mathfrak{s} \label{readresult}
\eeq
Thus, the shift due to curvature may be interpreted as a shift in the spin density relative to it's flat space value. Equation \eqref{readresult} is similar to the relation between Hall viscosity and spin presented in \cite{read2009,read2011}.

We expect that much of the physics that we have discussed in this paper will appear analogously in higher dimensional systems as well, although the details will be quite different. We will return to such cases elsewhere\cite{uslaterwork}.

\subsection*{Acknowledgments}
We thank E. Fradkin, S. Ryu, B. A. Bernevig, D. P. Arovas, P. Albin, D. Minic, S. Ramamurthy, A. Randono, O. Sule and A. Weiss for helpful discussions, and L. \'Alvarez-Gaum\'e and F. Bastianelli for email clarifications. Work supported by the US Department of Energy under contracts DE-FG02-07ER46453 (TLH) and FG02-91-ER40709 (RGL,OP).


\appendix\section{Appendix: Asymptotic expansions} \label{sec: AE}
\newcommand{\mtr}{\mathrm{Tr}}
\renewcommand{\theequation}{A.\arabic{equation}}
\setcounter{equation}{0}

On a number of occasions, we have encountered traces over Dirac fermions of the form
\beq
\mtr\;\gamma^5e^{\;t\slashed{\msD}^2},\;\mtr\;\gamma^5\xi^a\nabla_ae^{\;t\slashed{\msD}^2},\;\mtr\;\gamma^5\gamma^{ab}e^{\;t\slashed{\msD}^2},\;\mtr\;e^{\;t\slashed{\msD}^2} \label{traceeg}
\eeq
and in particular, their asymptotic expansions (in powers of \(t\)) in the limit \(t\rightarrow 0\). We can use \(\mathcal{N}=1\) supersymmetric quantum mechanics to evaluate these expressions.\footnote{The calculation of asymptotic expansions in presence of torsion has appeared in \cite{Obukhov1, NY, Kimura1}. The use of \(\mathcal{N}=1\) Supersymmetric quantum mechanics (SQM) in computing Chiral anomalies or Atiyah Singer index densities on torsional backgrounds has been discussed before in \cite{CZanelli2, waldron}, and in the special case of vanishing Nieh-Yan four form in \cite{Mavramatos, Kimura2, Bismut}.} We will not provide details, but rather only sketch the essential ideas involved; see \cite{AGW, BVN, deBoer1, deBoer2} for details.

Let \(\Sigma\) be a manifold with metric \(g_{ij}\), and a torsional connection \(\omega_{i;ab}=\lcw_{i;ab}+C_{i;ab}\). The action for \(\mathcal{N}=1\) SQM in the presence of torsion is given by
\beqn
S_{SQM} &=&\int dt\;\left(\frac{1}{2}g_{ij}\dot x^i\dot x^j+\frac{i}{2}\chi^a(\delta_{ab}\dot \chi^b+\dot x^k\lcw_{k;ab}\chi^b)-i\frac{\torcplg}{4}\dot x^k\chi^a\chi^b H_{kab}-\frac{\torcplg}{2}\frac{1}{4!}N_{abcd}\chi^a\chi^b\chi^c\chi^d\right.\nonumber\\
&+&\left. i\bar c(\dot c+i\dot x^kA_kc) + \frac{i}{2}\bar c F_{ab}\chi^a\chi^bc\right)
\eeqn
where \(x^i\) are local coordinates on \(\Sigma\), \(\chi^a\) are one-component real fermions, while \(c\) and \(\bar c\) are one-component complex fermions. The theory is invariant under the supersymmetry transformations \(\delta x^i = i\epsilon \chi^i,\;\delta \chi^i = -\epsilon \dot{x}^i\), with the supercharge
\beq 
Q = i\chi^a\underline{e}^i_a(p_i-\frac{i}{2}\lcw_{i,bc}\chi^b\chi^c+\bar cA_ic)-\frac{\torcplg}{2}\frac{1}{3!}H_{a;bc}\chi^a\chi^b\chi^c
\eeq
(\(p_i\) being the momentum conjugate to \(x^i\)), and the Hamiltonian \(\mathcal{H}=-Q^2\).
Upon quantization, we must replace \(p_i \rightarrow -i\partial_i\) and \(\chi^a \to \frac{1}{\sqrt{2}}\gamma^a\). The supercharge becomes \(Q = \frac{1}{\sqrt{2}}\slashed{\msD}+\cdots\), while the Hamiltonian is \(\mathcal{H} = -\frac{1}{2}\slashed{\msD}^2+\cdots\), upto operator ordering ambiguities indicated by \(\cdots\). Further, the fermion number operator in SQM, \((-1)^F\), is proportional to the chirality matrix \(\gamma^5\). This is what allows us to compute traces of the type \eqref{traceeg} - the Hilbert space of \(\mathcal{N}=1\) SQM essentially furnishes a representation of Dirac fermions on \(\Sigma\). 

For instance, consider \(\mtr\;\gamma^5e^{t\slashed{\msD}^2}\). In SQM, this is proportional to the Witten index \(\mtr\;(-1)^Fe^{-\beta\hat {\mathcal{H}}}\)
with \(t=\frac{1}{2}\beta\). Such a trace over the Hilbert space is easiest to compute using the path integral representation. To handle the operator ordering ambiguities, we follow the time-slicing prescription for the path integral \cite{BVN}, at the expense of the counterterms 
\beq
L_{ct}=\frac{1}{8}g^{ij}{{\mathring{\Gamma}}^k}_{\;\;il}{{\mathring{\Gamma}}^l}_{\;jk}+\frac{1}{16}{\omega^{(\torcplg)}}_{i;ab}{\omega^{(\torcplg)}}^{i;ab}-\frac{\torcplg^2}{16}\frac{1}{3!}H_{a;bc}H^{a;bc}
\eeq
The path integral corresponding to \(\mtr\;(-1)^Fe^{-\beta\hat{\mathcal{H}}}\) is then given by
\beq
\mtr\;(-1)^Fe^{-\beta\hat{\mathcal{H}}} = \int_{PBC} [dx^id\chi^ada^idb^idc^i]e^{-\int_{-\beta}^0dt\; L_E}
\eeq
where \(a_i\) are commuting ghosts, \(b_i\) and \(c_i\) are anti-commuting ghosts,\footnote{The ghosts are introduced to exponentiate factors of \(\mathrm{det}(e)\) which arise due to insertion of complete set of position eigenstates in the discretized path integral.} and \(L_E\) is the Euclidean time Lagrangian given by
\beqn
L_E  &=&\frac{1}{2}g_{ij}\dot x^i\dot x^j+\frac{1}{2}\delta_{ab}\chi^a\dot \chi^b+\frac{1}{2}\dot x^k\omega^{(\torcplg)}_{k;bc}\chi^b\chi^c+\frac{\torcplg}{2}N_{abcd}\chi^a\chi^b\chi^c\chi^d\nonumber\\
&+& \bar c(\dot c+\dot x^kA_k c) - \frac{i}{2}\bar c F_{ab}\chi^a\chi^b c+ \frac{1}{2}g_{ij}(a^ia^j+b^ic^j)+L_{ct}
\eeqn
Here \(x^i\) and \(a^i\) have periodic boundary conditions, \(\chi^a\) have periodic boundary conditions because of the \((-1)^F\) in the trace (which is what the subscript \(PBC\) indicates), and \(b_i,\; c_j,\;c\) and \(\bar c\) all have anti-periodic boundary conditions. In the absence of \((-1)^F\), \(\chi^a\) acquire anti-periodic boundary conditions \((APBC)\). Finally, the \(\beta \rightarrow 0\) limit is just the weak coupling limit in SQM, where we can do perturbation theory. In this way, \(\mathcal{N}=1\) SQM allows us to compute the asymptotic expansions in \eqref{traceeg} using standard techniques of field theory. For instance, using the method described above, we find the asymptotic expansion for \(\mtr\;\gamma^5e^{t\slashed{\msD}^2}\) in \(d=4\) is given by
\beqn
\mtr\;\gamma^5e^{t\slashed{\msD}^2}&\simeq&\int_{\Sigma}\left(\frac{\torcplg}{16\pi^2t} (T^a\wedge T_a - R_{ab}\wedge e^a\wedge e^b)+\frac{1}{8\pi^2}F\wedge F+\frac{1}{192\pi^2}{{R^{(-\torcplg)}}^a}_b\wedge {{R^{(-\torcplg)}}^b}_{a}\right.\nonumber\\
&+&\left.\frac{\torcplg}{96\pi^2}d\;d^{\dagger} (T^a\wedge T_a - R_{ab}\wedge e^a\wedge e^b)+ O(t)\right) \label{4dchiralanomaly}
\eeqn
The same procedure can be applied for computing the other asymptotic expansions in \eqref{traceeg}. We state some of the results relevant to the calculations of sections \ref{section 5} and \ref{sec: chiral gravity}. In \(d=2\) we have
\beqn
\mtr\;e^{t\slashed{\msD}^2} &\simeq& \int_{\Sigma}\left(\frac{1}{2\pi t}-\frac{1}{24\pi}\lcR+O(t)\right)vol_{\Sigma}\\
\mtr\;\gamma^5e^{t\slashed{\msD}^2}&\simeq& \int_{\Sigma}\left(\frac{1}{2\pi}F + O(t)\right)\\
\mtr\;\gamma^5\gamma^{ab}e^{t\slashed{\msD}^2}&\simeq&  \int_{\Sigma}-i\epsilon^{ab}\left(\frac{1}{2\pi t}-\frac{1}{24\pi}\lcR+O(t)\right)vol_{\Sigma}\label{lorentzanomaly}\\
\mtr\;\gamma^5\xi^a\nabla_ae^{-t\slashed{\msD}^2}&\simeq&\int_{\Sigma}-i\xi_a\left(\frac{1}{4\pi t}T^a+\frac{1}{48\pi}e^a\wedge d \lcR-\frac{1}{48\pi}\lcR T^a+O(t)\right)\label{diffanomaly}
\eeqn
while in \(d=3\) we need the result
\beq
\mtr\;e^{t\slashed{\msD}^2} \simeq \int_{\Sigma}\frac{2}{(4\pi t)^{3/2}}\left(1-\frac{t}{12\pi}R^{(-\torcplg)}+O(t^2)\right)vol_{\Sigma}
\eeq

\section{Appendix: Results and Conventions}
\renewcommand{\theequation}{B.\arabic{equation}}
\setcounter{equation}{0}

We work in a mostly plus metric $\eta_{ab}={\rm diag}(-1,1,\ldots, 1)$ and take the Clifford algebra to be
\beq
\{\gamma^a,\gamma^b\}=2\eta^{ab}
\eeq
represented by numerical unitary matrices.  $\gamma^0$ is then anti-Hermitian and all others are Hermitian. Furthermore, we have $\overline{\psi}=\psi^\dagger\gamma^0$. We use notation like $\gamma_{ab}=\frac{1}{2!} [\gamma_a,\gamma_b]$, and we have $\gamma^0\gamma^a(\gamma^0)^{-1}=-(\gamma^a)^\dagger$ and $(\gamma^0)^{-1}(\gamma^{ab})^\dagger\gamma^0=-\gamma^{ab}$, etc.
We note the following useful commutators and anti-commutators, valid in any dimension and signature
\beqn
\left\{\gamma^a,\gamma^{bc}\right\}&=&2\gamma^{abc} 
\\
 \left[\gamma^a,\gamma^{bc}\right]&=&2(\eta^{ab}\gamma^c-\eta^{ac}\gamma^b)\\
 \left\{\gamma^{ab},\gamma^{cd}\right\}&=& 2\gamma^{abcd}+2(\eta^{bc}\eta^{ad}-\eta^{bd}\eta^{ac})\\
\left[\gamma^{ab},\gamma^{cd}\right]&=& 2(\eta^{bc}\gamma^{ad}-\eta^{bd}\gamma^{ac}-\eta^{ac}\gamma^{bd}+\eta^{ad}\gamma^{bc})\\
\left\{\gamma^a,\gamma^{bcd}\right\}
&=&2(\eta^{ab}\gamma^{cd}+\eta^{ac}\gamma^{db}+\eta^{ad}\gamma^{bc})
\\
\left[\gamma^a,\gamma^{bcd}\right]&=&2\gamma^{abcd}
 \eeqn

%
%
%

\bibliographystyle{uiuchept}

\providecommand{\href}[2]{#2}\begingroup\raggedright\endgroup

\end{document}